\documentclass[12pt,a4paper, amsmath,amssymb]{article}
\usepackage[margin=0.8in]{geometry}
\usepackage{amssymb}
\usepackage{amsmath}
\usepackage{amsfonts}
\usepackage[utf8]{inputenc}
\usepackage{graphicx}
\usepackage{float}
\usepackage{xcolor}

\definecolor{bluemunsell}{rgb}{0.0, 0.5, 0.69}
\usepackage{float}
\usepackage{adjustbox}
\usepackage[colorlinks,linkcolor = bluemunsell,
urlcolor  = bluemunsell,
citecolor = bluemunsell,
anchorcolor = bluemunsell]{hyperref}\usepackage{hyperref}
\usepackage[normalem]{ulem}

\allowdisplaybreaks
\usepackage{xcolor}
\usepackage{makecell}
\usepackage{multirow}
\usepackage{tikz}

\usepackage{cite}
\usepackage{tikz}
\usepackage{ctable}
\usepackage{feynmp-auto}
\definecolor{green}{rgb}{0.0, 0.56, 0.0}

\providecommand{\openone}{\leavevmode\hbox{\small1\kern-3.8pt\normalsize1}}

\def\half{\tfrac{1}{2}}
\definecolor{green}{rgb}{0.0, 0.56, 0.0}
\usepackage{orcidlink}
\usepackage{etoolbox}
\makeatletter

\newcommand{\yu}{{y_{ij}^{u}}}
\newcommand{\yd}{{y_{ij}^{d}}}
\newcommand{\yt}{{y_{33}^{u}}}
\newcommand{\ytt}{{y_{34}^{u}}}

\newcommand{\yb}{{y_{33}^{d}}}

\newcommand{\Ql}{\overline{Q}_{Li}^0}
\newcommand{\Qlj}{\overline{Q}_{Li,j,4}^0}

\newcommand{\ur}{u_{Rj}^{0}}
\newcommand{\dr}{d_{Rj}^{0}}
\newcommand{\ult}{\bar{u}_{L3}^{0}}
\newcommand{\uli}{\bar{u}_{Li}^{0}}
\newcommand{\dli}{\bar{d}_{Li}^{0}}
\newcommand{\dlb}{\bar{d}_{L3}^{0}}
\newcommand{\urt}{{u}_{R3}^{0}}
\newcommand{\drb}{{d}_{R3}^{0}}
\newcommand{\T}{{u}_{R4}^{0}}
\newcommand{\slx}{s_L^u}
\newcommand{\slu}{s_L^u}
\newcommand{\sld}{s_L^d}
\newcommand{\clx}{c_L^u}
\newcommand{\clu}{c_L^u}
\newcommand{\cld}{c_L^d}

\newcommand{\sru}{s_R^u}
\newcommand{\srd}{s_R^d}

\newcommand{\cru}{c_R^u}
\newcommand{\crd}{c_R^d}

\newcommand{\uLm}{
	\begin{pmatrix}
		\bar{u}_{L3}^0 &\bar{u}_{L4}^0
		
	\end{pmatrix}
}

\newcommand{\dLm}{
	\begin{pmatrix}
		\bar{d}_{L3}^0 &\bar{d}_{L4}^0
		
	\end{pmatrix}
}

\newcommand{\uRm}{
	\begin{pmatrix}
		\bar{u}_{R3}^0 \\\bar{u}_{R4}^0
		
	\end{pmatrix}
}

\newcommand{\dRm}{
	\begin{pmatrix}
		{d}_{R3}^0 \\{d}_{R4}^0
		
	\end{pmatrix}
}
\newcommand{\Qlm}{
	\begin{pmatrix}
		\bar{u}_{L3}^0 &\bar{d}_{L3}^0
		
	\end{pmatrix}
}
\newcommand{\hun}{\frac{1}{\sqrt{2}}\Big[\cos\alpha h+\sin\alpha H-i\left(\sin\beta G^0+\cos\beta A\right)\Big]}

\newcommand{\hdc}{-\cos\beta G^-+\sin\beta H^-}

\newcommand{\hum}{
	\begin{pmatrix}
		\hun \\\huc
		
	\end{pmatrix}
}
\newcommand{\hdm}{
	\begin{pmatrix}
		\hdc \\\hdn
		
	\end{pmatrix}
}

\newcommand{\yuu}{
	\begin{pmatrix}
		{y_{33}^{u}}^* &{y_{34}^{u}}^*\\0&0		
	\end{pmatrix}
}

\newcommand{\uRmr}{
	\begin{pmatrix}
		{u}_{R3} \\{u}_{R4}
		
	\end{pmatrix}
}

\newcommand{\dRmr}{
	\begin{pmatrix}
		{d}_{R3} \\{d}_{R4}
		
	\end{pmatrix}
}

\newcommand{\uLmr}{
	\begin{pmatrix}
		\bar{u}_{L3} &\bar{u}_{L4}
		
	\end{pmatrix}
}

\newcommand{\dLmr}{
	\begin{pmatrix}
		\bar{d}_{L3} &\bar{d}_{L4}
		
	\end{pmatrix}
}
\makeatother
	
\begin{document}
		
		\begin{center}
			{\Large \textbf{Probing  Charged Higgs Bosons \\[0.25cm] in the 2-Higgs Doublet Model
					Type-II \\[0.45cm]  with Vector-Like Quarks
			}}
			\thispagestyle{empty}
			\def\thefootnote{\fnsymbol{footnote}}
			\vspace{1cm}
			
			{\sc
				R. Benbrik\orcidlink{0000-0002-5159-0325}$^1$\footnote{\href{mailto:r.benbrik@uca.ac.ma}{r.benbrik@uca.ac.ma}},
				M. Boukidi\orcidlink{0000-0001-9961-8772}$^1$\footnote{\href{mailto:mohammed.boukidi@ced.uca.ma}{mohammed.boukidi@ced.uca.ma}},
				S. Moretti\orcidlink{0000-0002-8601-7246}$^{2,3}$\footnote{\href{mailto:stefano.moretti@physics.uu.se}{stefano.moretti@physics.uu.se}; \href{mailto:s.moretti@soton.ac.uk}{s.moretti@soton.ac.uk}}\\}
	
			\vspace{1cm}
			{\sl\small
				$^1$Polydisciplinary Faculty, Laboratory of Fundamental and Applied Physics, Cadi Ayyad University, Sidi Bouzid, B.P. 4162, Safi, Morocco\\
				\vspace{0.1cm}
				$^2$Department of Physics \& Astronomy, Uppsala University, Box 516, SE-751 20 Uppsala, Sweden\\
				
				\vspace{0.1cm}
				$^3$School of Physics \& Astronomy, University of Southampton, Southampton, SO17 1BJ, United Kingdom
			}
		\end{center}
		\vspace*{0.1cm}
		\begin{abstract}
			{
		We study the phenomenology of charged Higgs bosons ($H^\pm$) and Vector-Like Quarks (VLQs), denoted as $T$, the latter possessing a charge identical to the top quark one, within the framework of the Two Higgs Doublet Model Type-II (2HDM-II). Upon examining two scenarios, one featuring a singlet $(T)$ (2HDM-II+$(T)$) and another a doublet $(TB)$ (2HDM-II+$(TB)$), we discover that the presence of VLQs has a significant effect on the (pseudo)scalar sector of the 2HDM-II. In particular, this leads to a reduction in the strict constraint on the mass of the charged Higgs boson, which is imposed by $B$-physics observables, specifically $B\to X_s\gamma$. {The observed reduction stems from modifications in the charged Higgs couplings to the Standard Model (SM) top and bottom quarks}. Notably, the degree of this reduction varies distinctly between the singlet 2HDM+$(T)$ and doublet 2HDM+$(TB)$ scenarios. Additionally, our investigation extends to constraints imposed by the oblique parameters $S$ and $T$ on the VLQ mixing angles. Furthermore, to facilitate efficient exploration of the '2HDM-II+VLQ' parameter space, we present results on pair production of VLQs $T$ ($pp\to T\bar T$), followed by $T\to H^\pm b$ and $H^\pm \to tb$ decays, yielding a distinctive $2t4b$ final state. This investigation thus provides valuable insights guiding the search for extended Higgs and quark sectors at the Large Hadron Collider (LHC) at CERN.}

		\end{abstract}
		
		\def\thefootnote{\arabic{footnote}}
		\setcounter{page}{0}
		\setcounter{footnote}{0}

		\newpage
		
		\section{Introduction}
		 
Evidence of either a light or heavy charged spin-0 boson, $H^\pm$, with a fundamental (i.e., point-like) structure would be a crucial piece of evidence suggesting physics Beyond the Standard Model (BSM), as such a state would not replicate any of those existing within the SM. The latter, in fact, includes charged bosons, but these are spin-1. Conversely, it includes a spin-0 boson, but this is chargeless. As a consequence, the search for charged Higgs bosons in recent years has received a great boost at existing accelerator machines like the Large
Hadron Collider (LHC) at CERN. However, despite all the efforts being made, still no $H^\pm$ signals have been seen. Thus, data from direct $H^\pm$ searches from both the ATLAS and CMS experiments have excluded large areas of the parameter spaces pertaining to BSM scenarios incorporating these new states. In particular, this has been done for a 2-Higgs Doublet Model Type-II (2HDM-II), including its Supersymmetric incarnation, the Minimal Supersymmetric Standard Model (MSSM). However, while the latter is able to escape LHCb indirect limits on $H^\pm$ states via, chiefly, $b\to s\gamma$ driven decays of $B$ mesons (thanks to new (s)particles in the corresponding loops cancelling the $H^\pm$ contributions), this is not possible for the former. Therefore, in the 2HDM-II, a rather stringent lower limit applies on its mass, $m_{H^\pm}$, of 580 GeV or so, quite irrespectively of the other model parameters.

Such a constraint can, however, be lifted if the 2HDM-II is supplemented with additional objects flowing inside the loops, enabling the $b\to s\gamma$ transitions, similarly to what happens in the MSSM. The purpose of this paper is to investigate whether a similar phenomenon can be triggered by the presence of Vector-Like Quarks (VLQs) alongside the particle states of the 2HDM-II.
VLQs are (heavy) spin-1/2 states that transform as triplets under colour, but, differently from the SM quarks, their Left- and Right-Handed (LH and RH) couplings have the same Electro-Weak (EW) quantum numbers. They are predicted by several theoretical constructs: e.g., models with a gauged flavour group \cite{Davidson:1987tr,Babu:1989rb,Grinstein:2010ve,Guadagnoli:2011id}, non-minimal
Supersymmetric scenarios \cite{Moroi:1991mg,Moroi:1992zk,Babu:2008ge,Martin:2009bg,Graham:2009gy,Martin:2010dc}, Grand Unified Theories  \cite{Rosner:1985hx,Robinett:1985dz}), little Higgs  \cite{Arkani-Hamed:2002ikv,Schmaltz:2005ky} and Composite Higgs  \cite{Dobrescu:1997nm,Chivukula:1998wd,He:2001fz,Hill:2002ap,Agashe:2004rs,Contino:2006qr,Barbieri:2007bh,Anastasiou:2009rv}
models, to name but a few. In most of these frameworks, VLQs appear as partners of the third generation of quarks, mixing with top and bottom quarks. Further, if their mass is not exceedingly large for the LHC kinematic reach, they can be accessible in its detectors in a variety of final states \cite{Aguilar-Saavedra:2009xmz,Okada:2012gy,DeSimone:2012fs,Buchkremer:2013bha,Aguilar-Saavedra:2013qpa}, so that some of these have already been explored by the ATLAS and CMS collaborations~\cite{ATLAS:2017vdo, ATLAS:2017nap, ATLAS:2018cye, CMS:2017ked,CMS:2017ynm,CMS:2018zkf,CMS:2017voh,CMS:2018kcw,ATLAS:2016ovj,ATLAS:2018mpo,ATLAS:2018uky,ATLAS:2018ziw,ATLAS:2018dyh,ATLAS:2021ddx,ATLAS:2021gfv,ATLAS:2021ibc, CMS:2016edj,CMS:2017fpk,CMS:2018haz,CMS:2018dcw,CMS:2018rxs,CMS:2020fuj,CMS:2020ttz,CMS:2022yxp,CMS:2022tdo,CMS:2022fck}.

Here, rather than invoking a complete theoretical scenario embedding VLQs (like, e.g., Compositeness), we adopt a simplified approach by manually extending the 2HDM-II with two VLQ representations (singlet and doublet), in the spirit of Refs.~\cite{Aguilar-Saavedra:2013qpa,Benbrik:2015fyz,Badziak:2015zez,Angelescu:2015uiz,Arhrib:2016rlj,Benbrik:2019zdp}, i.e., for the purpose of assessing the phenomenological consequences of such a modifications in general, irrespectively of the underlying theory. Indeed, we will be able to prove that the VLQ loops entering $b\to s\gamma$ transitions can cancel the large contributions due to $H^\pm$ ones, thereby reducing the aforementioned limit on $m_{H^\pm}$ down to approximately $200$ GeV for the doublet scenario (2HDM+$TB$) and to around 500 GeV in the singlet scenario(2HDM+$T$). {However, we shall see that achieving this reduction relies on large mixing angles, a condition largely contradicted by constraints imposed by the oblique parameters $S$ and $T$. The focus then will shift to the production of $H^\pm$ states through $pp\to T\bar T$ and $T\to H^\pm b$, with $H^\pm\to tb$, resulting in a final state characterised by two top and four bottom quarks ($2t4b$). }

The plan of this paper is as follows: The next section is devoted to introduce our 2HDM-II+VLQ framework. Then we proceed to discuss both theoretical and experimental bounds on it. Sect.~\ref{sec:results} presents our numerical results for the two chosen VLQ representations in turn, including providing Benchmark Points (BPs) amenable to further phenomenological investigation aimed at extracting signatures of the 2HDM-II+VLQ scenario at the LHC. Finally, we conclude. (We also have some appendices.)

\section{Model Descriptions}
In this section, we only provide a brief overview of the 2HDM-II+VLQ realisations that are relevant to our work.
Let us begin with recalling the well known $\mathcal{CP}$-conserving 2HDM scalar potential for two doublet fields $(\Phi_1, ~\Phi_2)$ with a discrete $\mathbb{Z}_2$
symmetry, $\Phi_1\to -\Phi_1$, that is only violated softly by dimension-2 terms \cite{Branco:2011iw,Gunion:1989we}:
\begin{eqnarray} \label{pot}
\mathcal{V} &=& m^2_{11}\Phi_1^\dagger\Phi_1+m^2_{22}\Phi_2^\dagger\Phi_2
-\left(m^2_{12}\Phi_1^\dagger\Phi_2+{\rm h.c.}\right)
+\half\lambda_1\left(\Phi_1^\dagger\Phi_1\right)^2
+\half\lambda_2\left(\Phi_2^\dagger\Phi_2\right)^2 \nonumber \\
&& \qquad +\lambda_3\Phi_1^\dagger\Phi_1\Phi_2^\dagger\Phi_2
+\lambda_4\Phi_1^\dagger\Phi_2\Phi_2^\dagger\Phi_1
+\left[\half\lambda_5\left(\Phi_1^\dagger\Phi_2\right)^2+{\rm h.c.}\right].
\end{eqnarray}
Here, all parameters are real. The two complex scalar doublets $\Phi_{1,2}$ may be rotated into a basis, $H_{1,2}$, where only one obtains a Vacuum Expectation Value (VEV). Using the minimisation conditions of the potential for the implementation of EW Symmetry Breaking (EWSB), the 2HDM can be fully described in terms of seven independent parameters: $m_h, m_H, m_A, m_{H^\pm},\tan\beta (=v_2/v_1), \sin(\beta - \alpha)$ and the soft breaking parameter $m^2_{12}$.
When we impose that no (significant) tree-level Flavour Changing Neutral Currents (FCNCs) are present in the theory, four Yukawa versions of the 2HDM can then be realised, depending on how the $\mathbb{Z}_2$ symmetry is implemented into the fermion sector. These are: Type-I, where only $\Phi_2$ couples to all fermions; Type-II, where $\Phi_2$ couples to up-type quarks and $\Phi_1$ couples to charged leptons and down-type quarks; Type-Y (or Flipped), where $\Phi_2$ couples to charged leptons and up-type quarks and $\Phi_1$ couples to down-type quarks; Type-X (or Lepton Specific), where $\Phi_2$ couples to quarks and $\Phi_1$ couples to charged leptons\footnote{In this paper, we will be discussing only Type-II.}.

We now move on to discuss the VLQ side of the model and  we  start by listing the representations of two gauge-covariant
multiplets ($(T)$ and $(TB)$) in Tab.~\ref{SM_rep}, where the fields $T$ and $B$ have electric
charges $2/3$ and $-1/3$, respectively. Specifically, the $T$ is chosen as triplet under the colour group $SU(3)_C$ and singlet under the EW group $SU(2)_L \times U(1)_Y$. Furthermore, the RH and LH components $T_{L,R}^0$  have the same EW and colour quantum numbers.
The mixing of VLQs with the first and second generations of SM quarks is heavily constrained by low energy physics measurement constraints.
One such example of such constraints originates from the EW Precision Observables (EWPOs), including oblique parameter corrections, these being radiative corrections to quantities such as the $W^\pm$ boson mass ($m_{W^\pm}$) and to the effective mixing angle $(\sin^2\theta_W)$ at high orders \cite{Peskin:1991sw, Abouabid:2023mbu}. Such corrections may be sensitive to VLQs and may impose restrictions on their masses and couplings. Additionally, as mentioned, VLQs, especially the third generation, may also affect the properties of top and bottom quarks through the mixing of fermions\cite{Altarelli:1990zd}. This may have implications for processes like the decay of the $Z$ boson into bottom quarks, which were measured with high precision at the LEP $e^+e^-$ collider at energies near the $Z$ resonance \cite{Vignaroli:2015ama, Boudjema:1989qga}.  Deviations from SM predictions in such measurements have the potential to provide strong constraints on the properties of VLQs. We thus focus on scenarios in which the VLQs interact solely to third-generation SM quarks.

\begin{table}[H]
	\centering
	{\renewcommand{\arraystretch}{1.25} 
		{\setlength{\tabcolsep}{1.25cm}
	\begin{tabular}{ccc}\hline  \hline
	Component fields	&$(T)$&	$(TB)$\\\hline 
		$U(1)_Y$&	2/3&	1/6  \\\hline 
		$SU(2)_L$&	1&	2 \\\hline 
		$SU(3)_C$&	3 &	3\\\hline\hline 
	\end{tabular} \nonumber 
			\caption{Singlet and doublet VLQ representations under the SM gauge group.}}}\label{SM_rep}
\end{table}
In the Higgs basis, the Yukawa Lagrangian can be written as:

\begin{equation}
-\mathcal{L} \,\, \supset  \,\, y^u \bar{Q}^0_L \tilde{H}_2 u^0_R +  y^d \bar{Q}^0_L H_1 d^0_R + M^0_u \bar{u}^0_L u^0_R  + M^0_d \bar{d}^0_L d^0_R + \rm {h.c}.
\end{equation}

Here, $u_R$ actually runs over $(u_R, c_R, t_R, T_R)$ and $d_R$ actually runs over $(d_R, s_R, b_R, B_R)$ while $y^{u,d}$ are 3$\times$4 Yukawa matrices.
When only the top quark “mixes” with $T$, the relation between mass eigenstates $(T_{L,R})$ and weak eigenstates $(T^0_{L,R})$ can be factored into two $2\times2$ unitary matrices $U_{L,R}$, such that
\begin{equation}
	\left(\! \begin{array}{c} t_{L,R} \\ T_{L,R} \end{array} \!\right) =
	U_{L,R}^u \left(\! \begin{array}{c} t^0_{L,R} \\ T^0_{L,R} \end{array} \!\right)
	= \left(\! \begin{array}{cc} \cos \theta_{L,R}^u & -\sin \theta_{L,R}^u e^{i \phi_u} \\ \sin \theta_{L,R}^u e^{-i \phi_u} & \cos \theta_{L,R}^u \end{array}
	\!\right)
	\left(\! \begin{array}{c} t^0_{L,R} \\ T^0_{L,R} \end{array} \!\right) \,,
	\label{ec:mixu}
\end{equation}
where $\theta$ is the mixing angle between mass and weak eigenstates and $\phi$ is a possible $\mathcal{CP}$-violating phase which will be ignored in our work. 
In the weak eigenstate basis, the diagonalisation of the mass matrices makes the Lagrangian of the third generation and heavy quark mass terms such that
\begin{eqnarray}
	\mathcal{L}_\text{mass} & = & - \left(\! \begin{array}{cc} \bar t_L^0 & \bar T_L^0 \end{array} \!\right)
	\left(\! \begin{array}{cc} y_{33}^u \frac{v}{\sqrt 2} & y_{34}^u \frac{v}{\sqrt 2} \\ y_{43}^u \frac{v}{\sqrt 2} & M^0 \end{array} \!\right)
	\left(\! \begin{array}{c} t^0_R \\ T^0_R \end{array}
	\!\right) \notag \\
	& & - \left(\! \begin{array}{cc} \bar b_L^0 & \bar B_L^0 \end{array} \!\right)
	\left(\! \begin{array}{cc} y_{33}^d \frac{v}{\sqrt 2} & y_{34}^d \frac{v}{\sqrt 2} \\ y_{43}^d \frac{v}{\sqrt 2} & M^0 \end{array} \!\right)
	\left(\! \begin{array}{c} b^0_R \\ B^0_R \end{array}
	\!\right) +\text{h.c.},
	\label{ec:Lmass}
\end{eqnarray}
where $M^0$ is a bare mass and the $y_{ij}$'s are Yukawa couplings. For the singlet case $y_{43}$ = 0, while for the doublet one has $y_{34}$ = 0.
Using  standard techniques of diagonalisation, the mixing matrices are obtained by
\begin{equation}
	U_L^q \, \mathcal{M}^q \, (U_R^q)^\dagger = \mathcal{M}^q_\text{diag} \,,
	\label{ec:diag}
\end{equation}
with $\mathcal{M}^q$ the two mass matrices in Eq.~(\ref{ec:Lmass}) and $\mathcal{M}^q_\text{diag}$ the diagonalised ones. 
The mixing angles in the LH and RH sectors are not independent parameters. Using
Eq.~(\ref{ec:diag}) and depending on the VLQs representation, one can find:
\begin{eqnarray}
	\tan \theta_R^q & = & \frac{m_q}{m_Q} \tan \theta_L^q \quad \text{(singlet)} \,, \notag \\
	\tan \theta_L^q & = & \frac{m_q}{m_Q} \tan \theta_R^q \quad \text{(doublet)} \,,
	\label{ec:rel-angle1}
\end{eqnarray}
with $(q,m_q,m_Q) = (u,m_t,m_T)$ and $(d,m_b,m_B)$.
		\section{Theoretical and Experimental Bounds}
		\label{sec-A}
In this section, we list the constraints that we have used to check the validity of our results.
From the theoretical side, we have the following requirements:
\begin{itemize}
	\item \textbf{Unitarity} constraints require the $S$-wave component of the various
	(pseudo)scalar-(pseudo)scalar, (pseudo)scalar-gauge boson, and gauge-gauge bosons scatterings to be unitary
	at high energy ~\cite{Kanemura:1993hm}.
	\item \textbf{Perturbativity} constraints impose the following condition on the quartic couplings of the scalar potential: $|\lambda_i|<8\pi$ ($i=1, ...5$)~\cite{Branco:2011iw}.    
	\item \textbf{Vacuum stability} constraints require the potential to be bounded from below and positive in any arbitrary
	direction in the field space, as a consequence, the $\lambda_i$ parameters should satisfy the conditions as~\cite{Barroso:2013awa,Deshpande:1977rw}:
	\begin{align}
	\lambda_1 > 0,\quad\lambda_2>0, \quad\lambda_3>-\sqrt{\lambda_1\lambda_2} ,\nonumber\\ \lambda_3+\lambda_4-|\lambda_5|>-\sqrt{\lambda_1\lambda_2}.\hspace{0.5cm}
	\end{align} 
\item \textbf{Constraints from EWPOs}, implemented through the oblique parameters\footnote{To compute the $S$ and $T$ parameters for the VLQ representations investigated in this study, we employed dimensional regularisation by employing the \texttt{FeynArts} and \texttt{FormCalc} packages \cite{Hahn:2000kx, Hahn:2001rv}.}, $S$ and $T$ ~\cite{Grimus:2007if},  require that, for a parameter point of our
model to be allowed, the corresponding $\chi^2(S^{\mathrm{2HDM\text{-}II}}+S^{\mathrm{VLQ}},~T^{\mathrm{2HDM\text{-}II}}+T^{\mathrm{VLQ}})$ is within 95\% Confidence Level (CL) in matching the global fit results \cite{Molewski:2021ogs}:
\begin{align}
S= 0.05 \pm 0.08,\quad T = 0.09 \pm 0.07,\nonumber\quad  \rho_{S,T} = 0.92 \pm 0.11.\hspace{1.5cm} 
\end{align}
Note that unitarity, perturbativity, vacuum stability, as well as $S$ and $T$ constraints, are enforced through the public code  \texttt{2HDMC-1.8.0}\footnote{The code has been adjusted to include new VLQ couplings, along with the integration of analytical expressions for $S_{VLQs}$ and $T_{VLQs}$ outlined in Appendix \ref{appSTU}.} \cite{Eriksson:2009ws}.
\end{itemize}
From the experimental side, we evaluated the following:
\begin{itemize}
\item \textbf{Constraints from the SM-like Higgs-boson properties}  are taken into account by using \texttt{HiggsSignal-3} \cite{Bechtle:2020pkv,Bechtle:2020uwn} via \texttt{HiggsTools} \cite{Bahl:2022igd}. We require that the relevant quantities (signal strengths, etc.) satisfy $\Delta\chi^2=\chi^2-\chi^2_{\mathrm{min}}$ for these measurements at 95\% CL ($\Delta\chi^2\le6.18$).
\item\textbf{Constraints from direct searches at colliders}, i.e., LEP, Tevatron, and LHC, are taken at the 95\% CL and are tested using \texttt{HiggsBouns-6}\cite{Bechtle:2008jh,Bechtle:2011sb,Bechtle:2013wla,Bechtle:2015pma} via \texttt{HiggsTools}. Including the most recent searches for neutral and charged scalars.

\item {\bf Constraints from flavour physics} are taken at 95\% CL from experimental measurements are accounted for via the public code \texttt{SuperIso\_v4.1} \cite{Mahmoudi:2008tp}. Specifically, we have used the following measurements for the most relevant Branching Ratios ${\cal BR}$s:
	\begin{enumerate}
		\item ${\cal BR}(\overline{B}\to X_s\gamma)|_{E_\gamma<1.6\mathrm{~GeV}}=\left(3.32\pm0.15\right)\times 10^{-4}$~\cite{HFLAV:2016hnz},
		\item ${\cal BR}(B^+\to \tau^+\nu_\tau)=\left(1.06\pm0.19\right)\times 10^{-4}$~\cite{HFLAV:2016hnz},
		\item  ${\cal BR}(B_s\to \mu^+\mu^-) =\left(3.83^{+0.38}_{-0.36}\right)\times 10^{-9}$~\cite{CMS:2022mgd}
		\item  ${\cal BR}(B^0\to \mu^+\mu^-)=\left(1.2^{+0.8}_{-0.7}\right)\times 10^{-10}$~\cite{LHCb:2021awg,LHCb:2021vsc}
		
	\end{enumerate} 

\end{itemize}
		

		\section{Numerical Results}\label{sec:results}
In this section, we present our numerical results by first describing the Lagrangian terms relevant to  $b\to s\gamma$ transitions in the 2HDM-II+VLQ scenarios considered, then by charactering the implications of the latter in the case of the $(T)$ and $(TB)$ representations chosen here in turn, and, finally, by presenting some BPs amenable to experimental investigation.
\subsection{Charged Higgs Boson Contributions to Flavour Observables}
In the following, we shall discuss the results of the most relevant $B$-physics constraints, from $b\to s\gamma$ transitions, onto the 2HDM-II+VLQ scenarios considered.

The Lagrange density, which defines the interactions of the charged Higgs boson with third generation fermions can be written as:
\begin{eqnarray}
-\mathcal{L}_{H^+} ~~&=&    \frac{\sqrt{2}}{v}\overline{t}\left(\kappa_{t}m_{t}P_L-\kappa_{b}m_{b}P_R\right)bH^++{\rm h.c.},
\end{eqnarray}
where $P_{L/R} = (1\pm \gamma^5)/2$ are the chiral projection operators. For the 2HDM-II, 2HDM-II+$(T)$, and 2HDM-II+$(TB)$ cases, the couplings $\kappa_{t}$ and $\kappa_{b}$ take the values presented in Tab. \ref{coulping}. We emphasise here the role of the Yukawa couplings, as it is the changes of the latter occurring in presence of VLQs which are responsible for the forthcoming results, as opposed to the role of the VLQs in the loop observables that we will be describing, owing to the far too large values of their masses (of order 750 GeV or more).

	\begin{table}[H]
	\centering
{\renewcommand{\arraystretch}{1.4} 
	{\setlength{\tabcolsep}{0.05cm}
				\begin{tabular}{c||c||c}
					\hline\hline
					Models& $\kappa_{t}$  &$\kappa_{b}$	\\	\hline\hline
					{2HDM-II}&$\cot\beta$ &$-\tan\beta$\\\hline
					{2HDM-II+}${(T)}$&$c_L\cot\beta$&$-c_L\tan\beta$\\\hline
					{2HDM-II+}$(TB)$& $\cot\beta\left[c_L^d c_L^u + \frac{s_L^d}{s_L^u } (s_L^u{}^2 - s_R^u{}^2) e^{i(\phi_u - \phi_d)}\right]$   & $-\tan\beta\left[ c_L^u c_L^d + \frac{s_L^u}{s_L^d } (s_L^d{}^2 - s_R^d{}^2) e^{i(\phi_u - \phi_d)} \right]$ \\\hline\hline
		\end{tabular}}}
		\caption{Yukawa couplings of the  charged Higgs bosons $H^\pm$  to the third generation of quarks in the 2HDM-II and 2HDM-II+VLQ representations studied here ($(T)$ and $(TB)$).  Here $s_{L,R} = \sin \theta_{L,R}$ and 
			$c_{L,R} =\cos\theta_{L,R}$.}\label{coulping}
	\end{table}

The table summarises the Yukawa couplings of charged Higgs bosons to quarks of the third generation for three distinct representations: 2HDM-II, 2HDM-II+$(T)$, and 2HDM-II+$(TB)$, providing insights into the modification patterns in different scenarios.
As indicated, the couplings of the charged Higgs boson to SM quarks, specifically to the top ($t$) and bottom ($b$), undergo modifications in both VLQ representations\footnote{For the detailed analytic calculation of charged Higgs boson couplings, please refer to Appendix A.}. These alterations lead to significant changes in observables related to $B$-physics processes, such as $b \to s\gamma$ and $B_{s/d} \to \mu^+\mu^-$. The contributions to the corresponding Wilson coefficients ($C_{7,8}$) are proportional to $\kappa_i\kappa_j^*$, where the terms can be further decomposed into two parts. Detailed expressions of $C_{i,\kappa_{b}\kappa_{t}^*}^{t}$ can be found in Ref.~\cite{Hermann:2012fc}.

The contributions\footnote{In this study, we neglect contributions from Feynman diagrams involving VLQs, focusing solely on diagrams that involve SM quarks. This deliberate exclusion stems from the observation that the impact of VLQ diagrams is negligible when contrasted with the diagrams incorporating SM quarks.} to $C_i^{t,\mathrm{model}}$ take the form:

\begin{equation}
C_i^{t,\mathrm{model}} = \kappa_b\kappa_{t}^*C_{i,\kappa_{b}\kappa_{t}^*}^{t,\mathrm{model}} + \kappa_t\kappa_{t}^*C_{i,\kappa_{t}\kappa_{t}^*}^{t,\mathrm{model}} \label{wilson}
\end{equation}

The outcomes of these modifications will be presented in the following subsections.

\subsection{2HDM-II+${(T)}$}

\begin{figure}[H]
	\centering
	\includegraphics[width=1.\textwidth,height=0.475\textwidth]{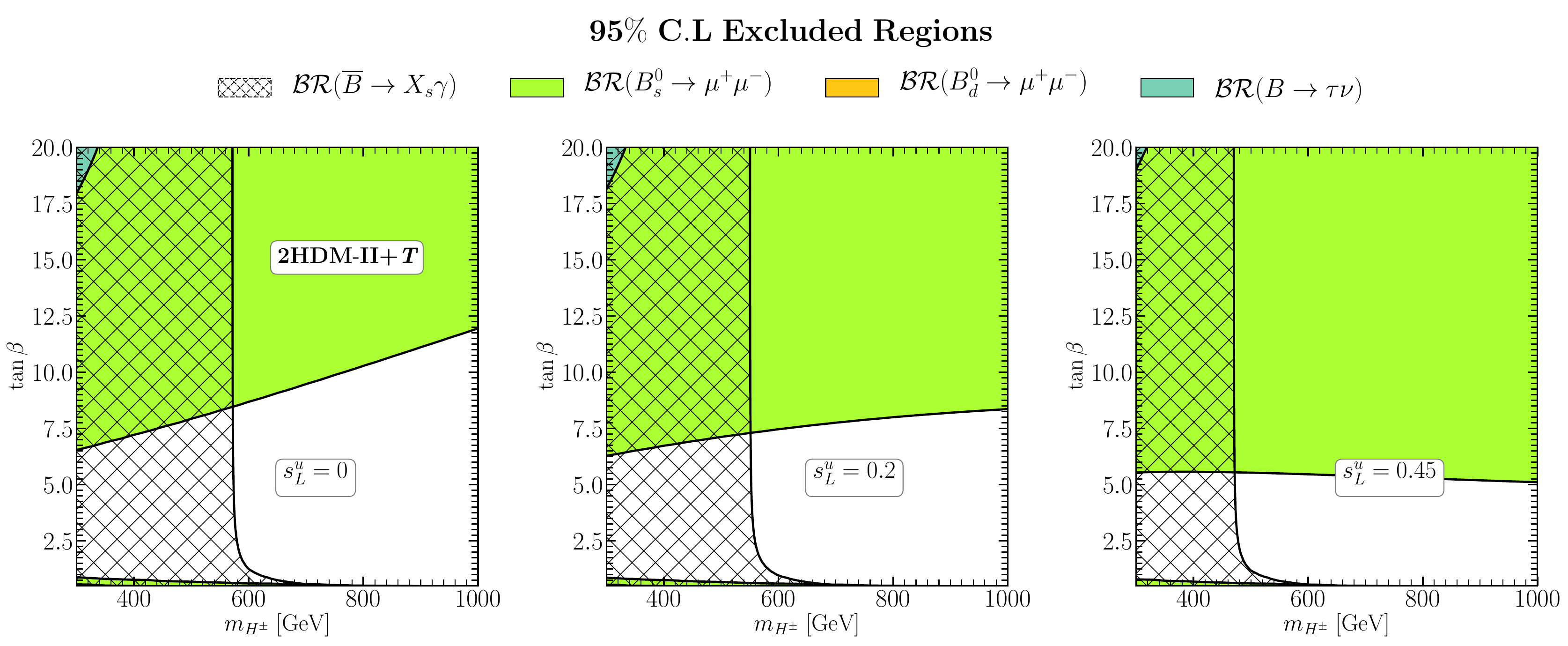}
	\caption{Excluded regions of the $(m_{H^\pm} , \tan\beta)$ parameter space by flavour constraints at 95\%
		CL.  Plots are presented for the 2HDM-II+$(T)$ singlet  with $s_L^u=0$ (left),  $s_L^u=0.2$ (middle) and $s_L^u=0.45$ (right).}
	\label{fig1}
\end{figure}

We start  with Fig.~\ref{fig1}, where we present explicitly the excluded parts of the ($m_{H^\pm}, \tan\beta$) plane at 95\%  CL by $\bar{B}\to X_{s}\gamma$ (hatched areas) alongside  $B_{d}^{0}\to \mu^{+}\mu^{-}$ (green), $B_{s}^{0}\to \mu^{+}\mu^{-}$ (orange, which is hardly visible in the plots),   and $B_{u}\to \tau\nu$ (blue). This figure illustrates that higher values of $s_L$ lead to a less stringent constraint from $B\to X_s \gamma$, driving the exclusion regions towards smaller values of the charged Higgs boson mass, reaching approximately $m_{H^\pm}\simeq500$ GeV for $s_L=0.45$. Conversely, the limits from $B_s\to \mu^+\mu^-$ become more restrictive as $s_L$ increases, excluding $\tan\beta\lesssim$ 7 for $s_L=0.25$ and $\tan\beta\lesssim 5$ for $s_L=0.45$. Here,  it is important to note that the new term introduced in the relevant $H^\pm tb$ coupling depends solely on the mixing angle $s_L$, which thus controls  the inclusion of VLQs in the 2HDM-II, as discussed  previously. 

However, the potential for this reduction is moderated by the oblique parameters $S$ and $T$, as will be discussed subsequently. These parameters restrict the range of the mixing angle $s_L$, thereby playing a critical role in preventing a substantial decrease in the charged Higgs mass limit, particularly by constraining the possibility of larger mixing angles.
		\begin{figure}[H]	
			\centering
			\includegraphics[height=9cm,width=10cm]{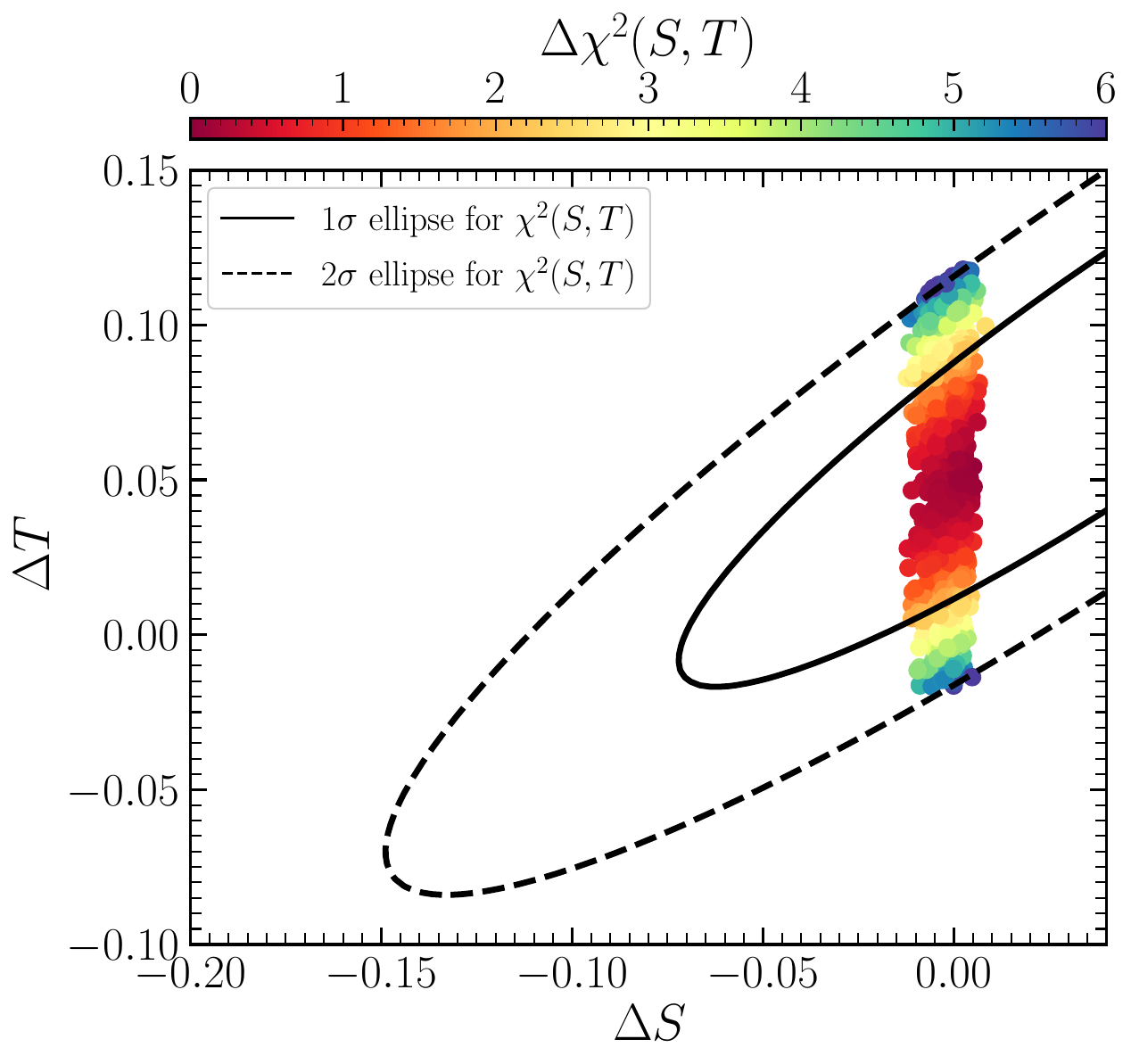}
			\caption{Allowed points by all constraints (\texttt{HiggsBounds}, \texttt{HiggsSignals}, \texttt{SuperIso} and theoretical ones) superimposed onto the fit limits in the
				$( S, ~ T)$ plane from EWPO data at 95\% CL (with a correlation of 92\%), with the colour
				code indicating $\Delta\chi^2(S,T)$. Results are presented for 
				the 2HDM-II+$(T)$.}\label{fig2}
		\end{figure}
	
	 		\begin{figure}[H]
		\centering
	\includegraphics[height=8cm,width=10cm]{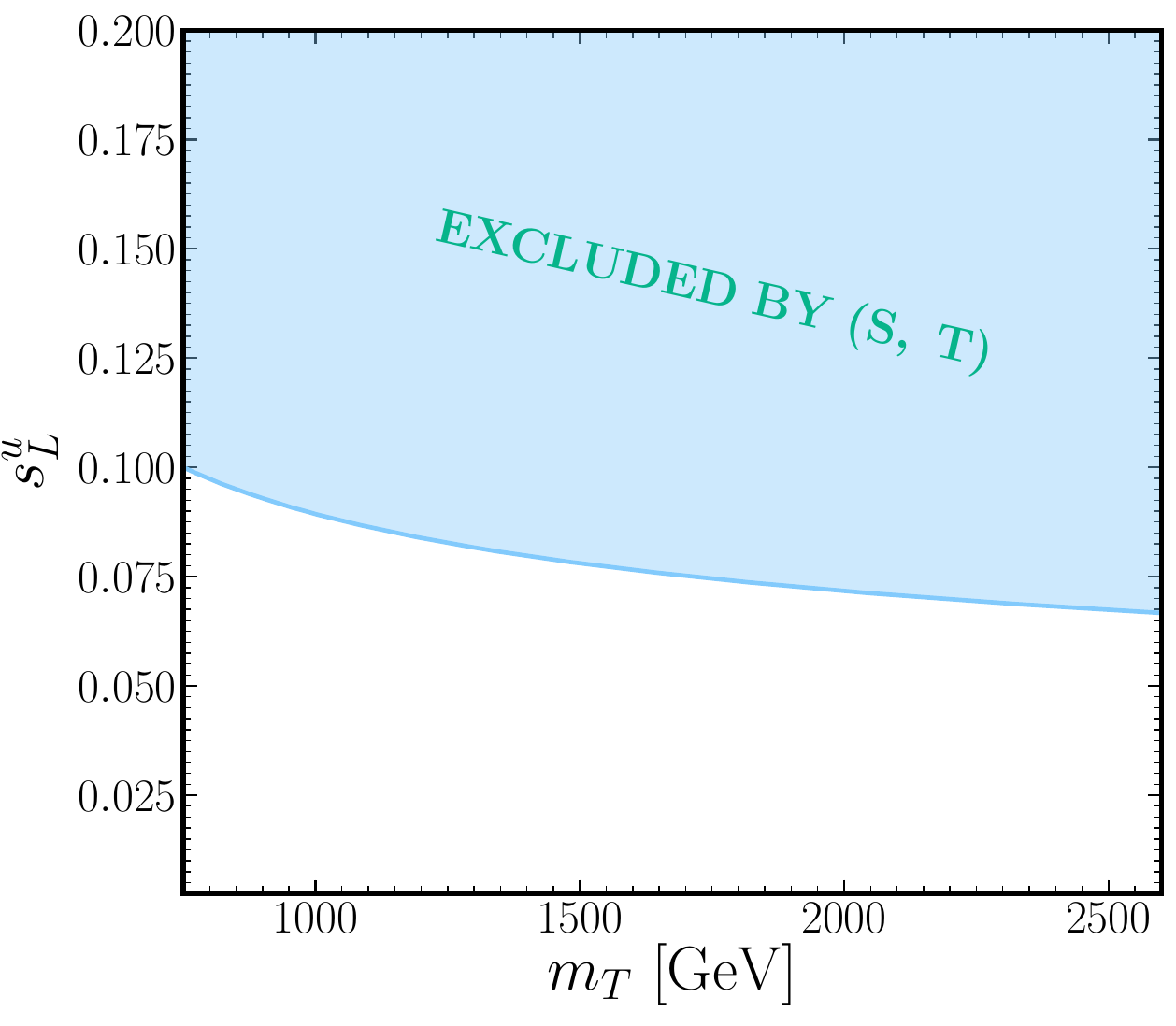}
		
		\caption{Exclusion regions by $(S,~T)$ in the $(m_T,~s^d_R)$ plane with fixed parameters: $m_{H}=532.80$ GeV, $m_A=524.26$ GeV and $m_{H^\pm}=588.02$ GeV, with $\tan\beta=5.19$.}\label{fig22}
	\end{figure}

\begin{table}[H]
	\centering
	{\renewcommand{\arraystretch}{1.0} 
		{\setlength{\tabcolsep}{2.75cm}
			\begin{tabular}{c  c}
				\hline\hline
				Parameters  & Scanned ranges \\
				\hline\hline
				
				\multicolumn{2}{c}{2HDM} \\\hline\hline						
				$m_H$  & [130, 800] \\
				$m_A$  & [80, 800] \\
				$m_{H^\pm}$  & [80, 800] \\
				$\tan\beta$ & [0.5, 20] \\
				$\sin(\beta-\alpha)$ & 1 \\
				\hline\hline
				\multicolumn{2}{c}{ 2HDM-II+$(T)$ } \\\hline\hline
				$s_L$  & [$-$0.5, 0.5] \\
				$m_{T}$   & [750, 2600] 	
				\\\hline\hline
				\multicolumn{2}{c}{2HDM-II+$(TB)$ } \\\hline\hline
					$s_{R}^{u,d}$  & [$-$0.5, 0.5] \\
				     $m_{T}$   & [750, 2600] \\	
				\hline\hline
	\end{tabular}}}
	\caption{2HDM and VLQs parameters with their scanned ranges for both VLQ representations considered here. Masses are in GeV. (Note that we have taken the lightest  $\mathcal{CP}$-even neutral Higgs boson, $h$, to be the one observed at $ \sim$ 125 GeV.) }
	\label{parameters}

\end{table}
In Fig.~\ref{fig2}, we show the result of the analysis for our scan over the $(S,T)$ plane. The colour coding indicates the difference with respect to the $\chi^2(S,T)$ values, and the black lines show the 68\% CL (solid) and 95\% CL (dashed) contours. The scanned 2HDM-II+VLQ parameters\footnote{We performed a systematic random scan over the parameters cited in Tab.~\ref{parameters} using a modified version of \texttt{2HDMC-1.8.0}.} are given in Tab.~\ref{parameters}. In the remainder of our analysis, we will present results from the 2HDM-II+($T$) regions within the $2\sigma$ band also compliant with flavour constraints. It is important to highlight that the shape of these points is a result of the cancellation between the contributions from the 2HDM and VLQ states to the oblique parameters $S$ and $T$.

{Fig.~\ref{fig22} illustrates the excluded regions at 95\%  CL by $S$ and $T$ in the ($m_T$, $s^u_L$) plane, considering the fixed parameters $m_{H}=532.80$ GeV, $m_A=524.26$ GeV, and $m_{H^\pm}=588.02$, with $\tan\beta=5.19$. From this figure, it is evident that the imposed constraints only permit low mixing angles, specifically $s^u_L\lesssim 0.1$. Additionally, the stringency of the constraint increases with rising VLQ mass, $m_T$.}

\begin{figure}[H]
	\begin{minipage}{0.46\textwidth}
		\centering
		\includegraphics[height=9cm,width=8.3cm]{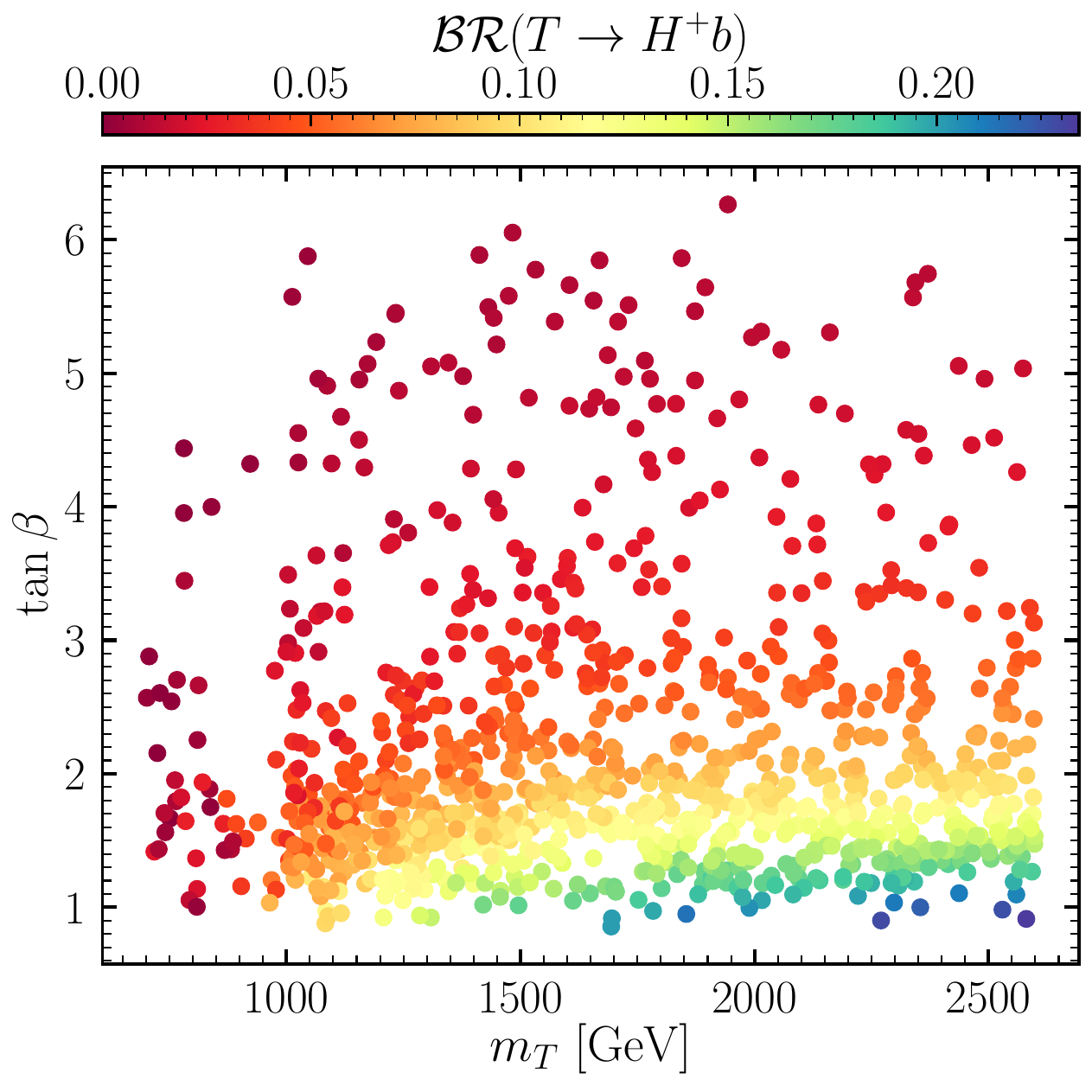}
	\end{minipage}\hspace{0.5cm}
	\begin{minipage}{0.46\textwidth}
		\centering
		\includegraphics[height=9cm,width=8.3cm]{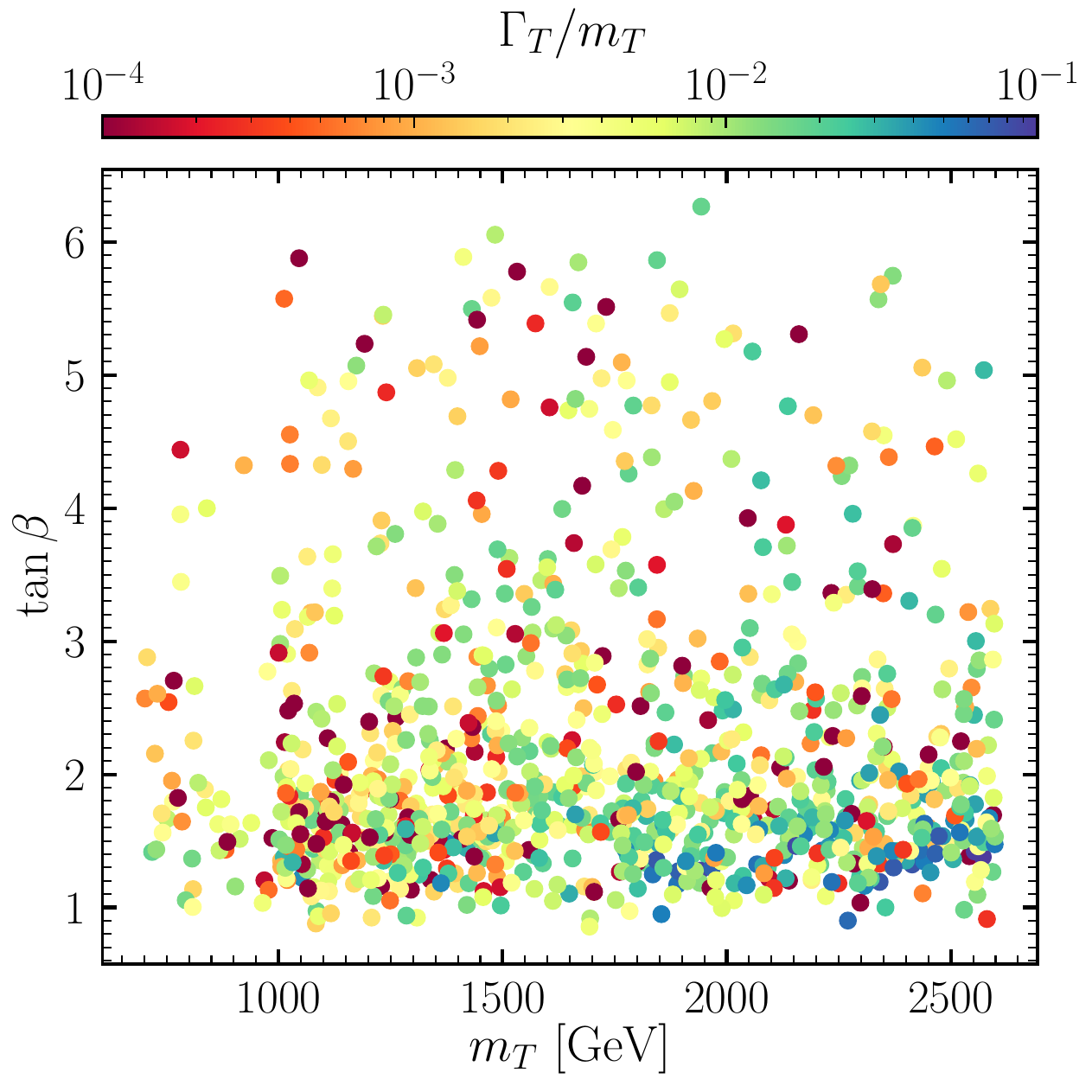}
	\end{minipage}			
	\caption{$\mathcal{BR}(T\to H^+ b)$ (left) and  $\Gamma_T/m_T$ (right) plotted over the $(m_T,~\tan\beta)$ plane. Results are presented for 
				the 2HDM-II+$(T)$.}\label{fig3}
\end{figure}

In Fig.~\ref{fig3}, we illustrate the  $\mathcal{BR}(T\to H^+ b)$ (left) and the ratio $\Gamma_T/m_T$ (right) as a function of $m_T$ and $\tan\beta$.
It is clear from the left plot that the 2HDM-II+$(T)$ scenario cannot ultimately predict dominant production rates for a $H^\pm$ emerging from a heavy $T$ quark, no matter its final decay pattern. Specifically,
the decay rate of $T\to H^\pm b$ reaches a maximum value of {23\%  for $\tan\beta\le1.5$ and $m_T\ge 1500$ GeV} and becomes negligible for $\tan\beta\ge4$, essentially due to the dependence of the $\mathcal{BR}$ on the mixing angle $s_L$. Recall, in fact, that the latter is constrained to be rather small by the oblique parameters $S$ and $T$. One can also read from the right panel of the figure that the heavy quark total width can be at {most 10\% of its mass at low $\tan\beta$ (indicated by dark blue points)}, so that this state is generally rather narrow and then can (tentatively, depending on the $H^\pm$ decays) be reconstructed in LHC analysis.

\begin{figure}[H]
	\begin{minipage}{0.46\textwidth}
	\centering
	\includegraphics[height=9cm,width=8.3cm]{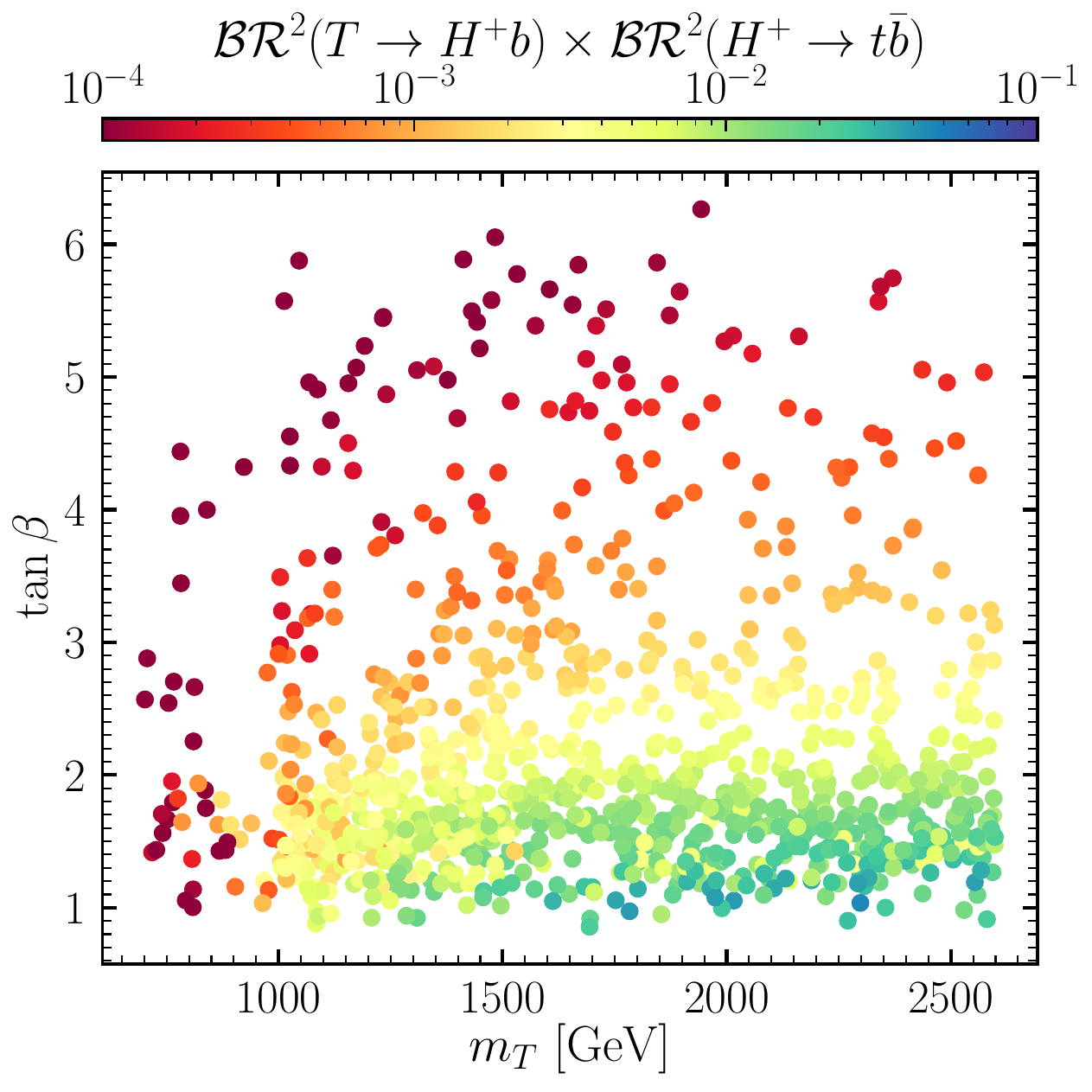}
\end{minipage}\hspace{0.5cm}	
\begin{minipage}{0.46\textwidth}
	\centering
	\includegraphics[height=9cm,width=8.3cm]{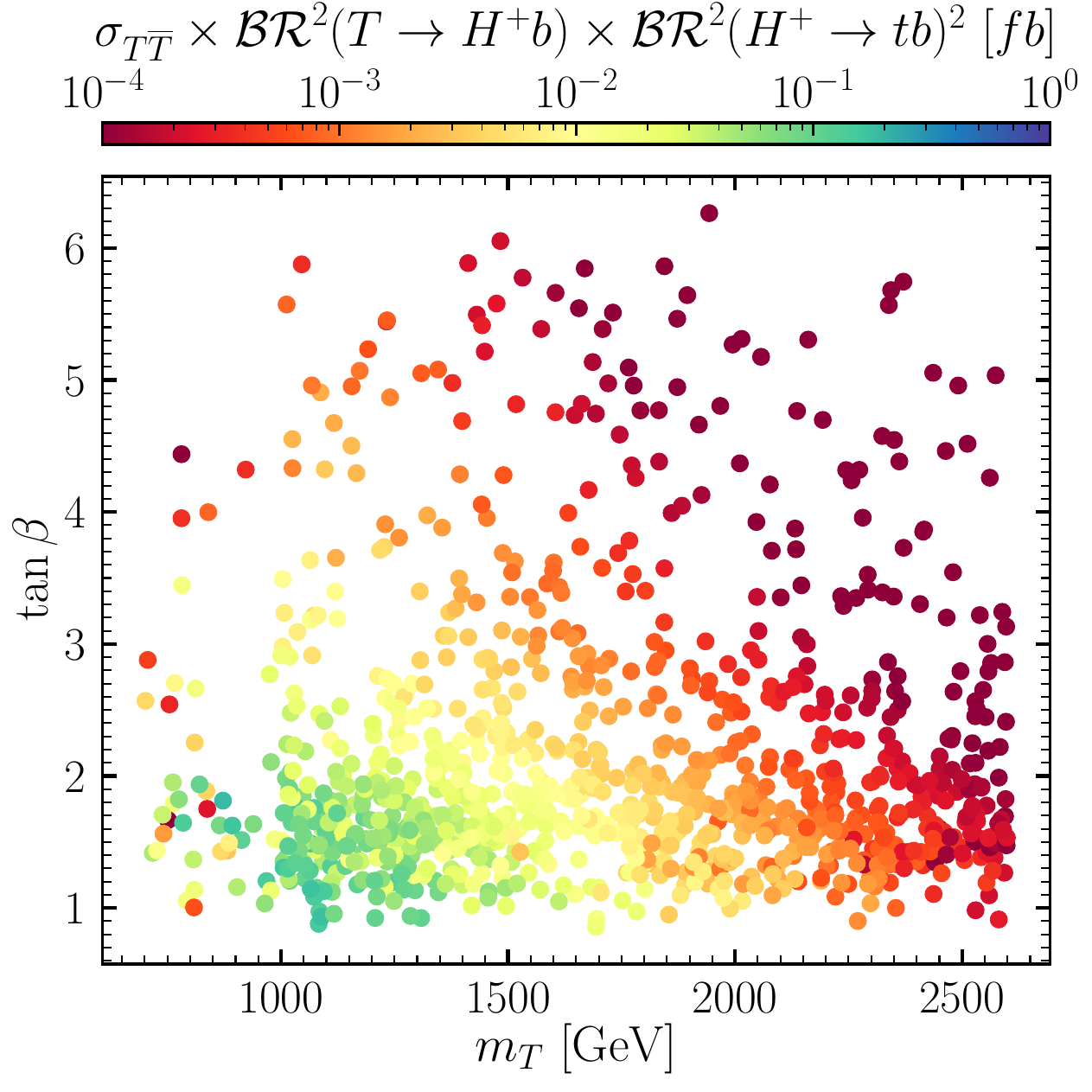}
\end{minipage}

\caption{$\mathcal{BR}^2(T\to H^+ b)\times\mathcal{BR}^2(H^+\to tb)$ (left) and  $\sigma_{T\overline{T}}\times\mathcal{BR}^2(T\to H^+ b)\times\mathcal{BR}^2(H^+\to tb)$ (right) plotted over the $(m_T,~\tan\beta)$ plane. Results are presented for the 2HDM-II+$(T)$. The two selected BPs are highlighted in green stars (see later).}\label{fig5}
\end{figure}

We then present in Fig.~\ref{fig5} the $\mathcal{BR}^2(T\to H^+ b)\times\mathcal{BR}^2(H^+\to t\bar{b})$ (left) and the production cross section $\sigma(pp\to T\overline{T})$\footnote{The cross section $\sigma(pp\to T\overline{T})$ is computed at LO using \texttt{MadGraph5}\_\texttt{aMC}{\fontfamily{pag}\fontsize{11}{1}\selectfont@}\texttt{NLO-3.4.0} \cite{Alwall:2014hca} for $\sqrt{s}$= 14 TeV, with \texttt{CTEQ6L} \cite{Pumplin:2002vw} as the Parton Distribution Functions (PDFs) with default settings.} followed by $\mathcal{BR}^2(T\to H^+ b)$ and $\mathcal{BR}^2(H^+\to t\bar{b})$ in the ($m_T, m_{H^\pm}$) plane. Clearly from the left panel, the decay $H^+\to t\bar{b}$ as this is the dominant one for a heavy $H^\pm$ state with $m_{H^\pm}>m_t$.
In the left panel, $\mathcal{BR}^2(T\to H^+ b)\times\mathcal{BR}^2(H^+\to t\bar{b})$ reaches around {4\% for $\tan\beta\le2$} while, in the right panel,  $\sigma_{T\overline{T}}\times\mathcal{BR}^2(T\to H^+ b)\times\mathcal{BR}^2(H^+\to t\bar{b})$ can exceed {0.2 fb} for $m_T<1000$ GeV and for small $\tan\beta$. The emerging $2t4b$ signal\footnote{{A similar signal has been investigated within the 2HDM-II+VLQs framework in \cite{Dermisek:2020gbr}.}} can then potentially be pursued at the LHC \cite{Dermisek:2020gbr}.


		\subsection{2HDM-II+${(TB)}$}
In this section, we discuss the case of the 2HDM-II+$(TB)$.  In the SM extended with such a VLQ multiplet, both mixing angles in the up- and down-type quark sectors enter the phenomenology of the model. For a given $\theta_R^b$, $\theta_R^t$, and $m_T$ mass, the relationship between the mass eigenstates and the mixing angles is given by \cite{Aguilar-Saavedra:2013qpa}:
\begin{eqnarray}
	&&m_B^2= (m_T^2 \cos^2\theta_R^{t}+m_t^2 \sin^2\theta_R^{t}-
	m_b^2 \sin^2\theta_R^{b})/ \cos^2\theta_R^{b}. 
	\label{tb-mass}
\end{eqnarray}
Using the above relation, one can then compute $m_B$ for a given $m_T$ and mixing angles $\theta_R^b$ and $\theta_R^t$. Additionally, the left mixings $\theta_L^b$, $\theta_L^t$ may be calculated using Eq.~(\ref{ec:rel-angle1}).
\begin{figure}[H]
	\centering
	\includegraphics[width=1.\textwidth,height=0.475\textwidth]{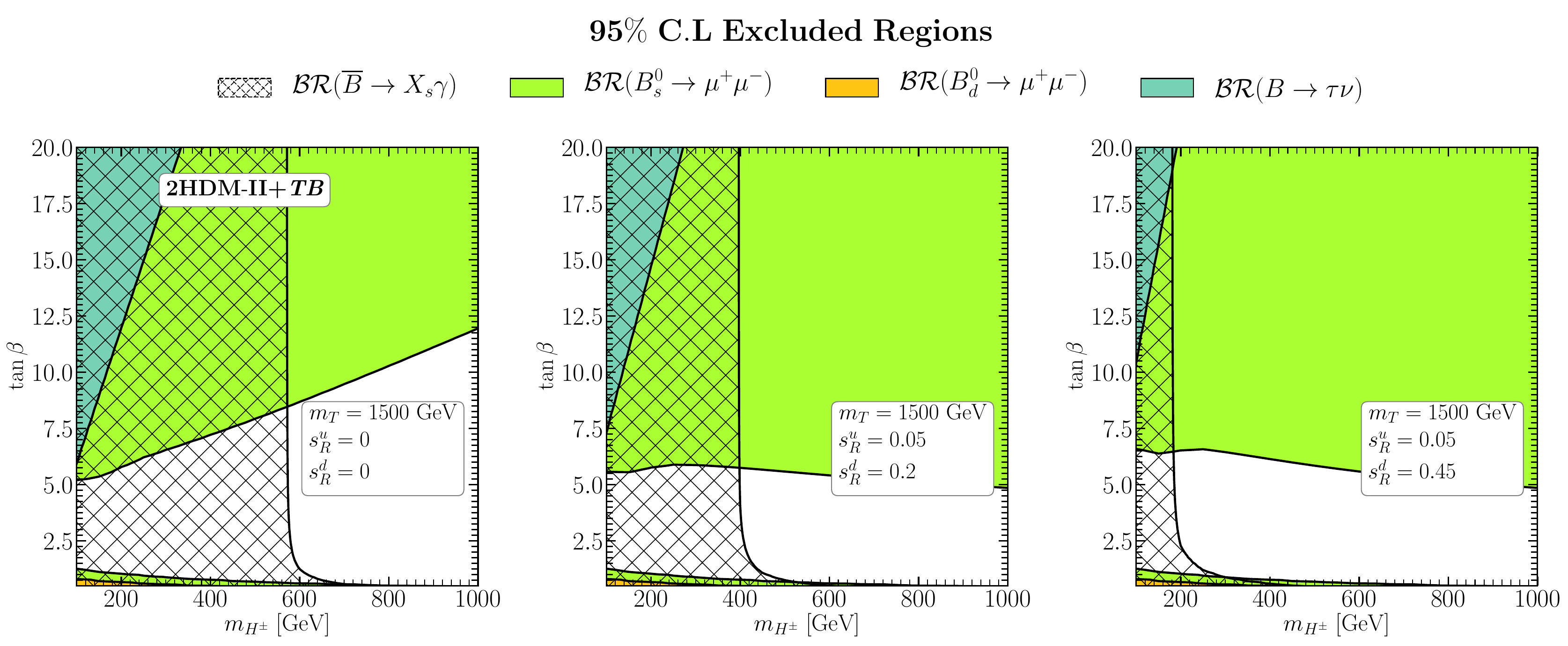}
	\caption{Excluded regions of the $(m_{H^\pm} , \tan\beta)$ parameter space by flavour constraints at 95\%
		CL.  Plots are presented for the 2HDM-II+$(TB)$ doublet  with $s^d_R=0$ (left),  $s^d_R=0.2$ (middle) and $s^d_R=0.45$ (right), with fixed $s_R^u=0.05$.}
	\label{s^d_R}
\end{figure}

In Fig.~\ref{s^d_R}, we once again present the excluded regions over the ($m_{H^\pm}, \tan\beta$) plane at a 95\% CL. The exclusions are derived from $\bar{B}\to X_{s}\gamma$ (depicted as hatched areas) as well as from $B_{d}^{0}\to \mu^{+}\mu^{-}$ (depicted in green), $B_{s}^{0}\to \mu^{+}\mu^{-}$ (depicted in orange), and $B_{u}\to \tau\nu$ (depicted in blue).
Notably, in this representation, the limit imposed by $\bar{B}\to X_{s}\gamma$ can be further diminished compared to the singlet scenario. The exclusions shift the regions for $B\to X_s \gamma$ towards smaller values of the charged Higgs boson mass, particularly reaching $m_{H^\pm}\sim 400$ GeV for $s_R^d=0.2$ and $m_{H^\pm}\simeq180$ GeV for $s_R^d=0.45$, maintaining a fixed value of $s^u_R$ at 0.05 in both cases. Similarly to the previous scenario, the constraints from $B_s\to\mu^+\mu^-$ become more stringent than in the 2HDM case in this representation, excluding all values above $\tan\beta\simeq5$. 

\begin{figure}[H]	
	\centering
	\includegraphics[height=8.8cm,width=10cm]{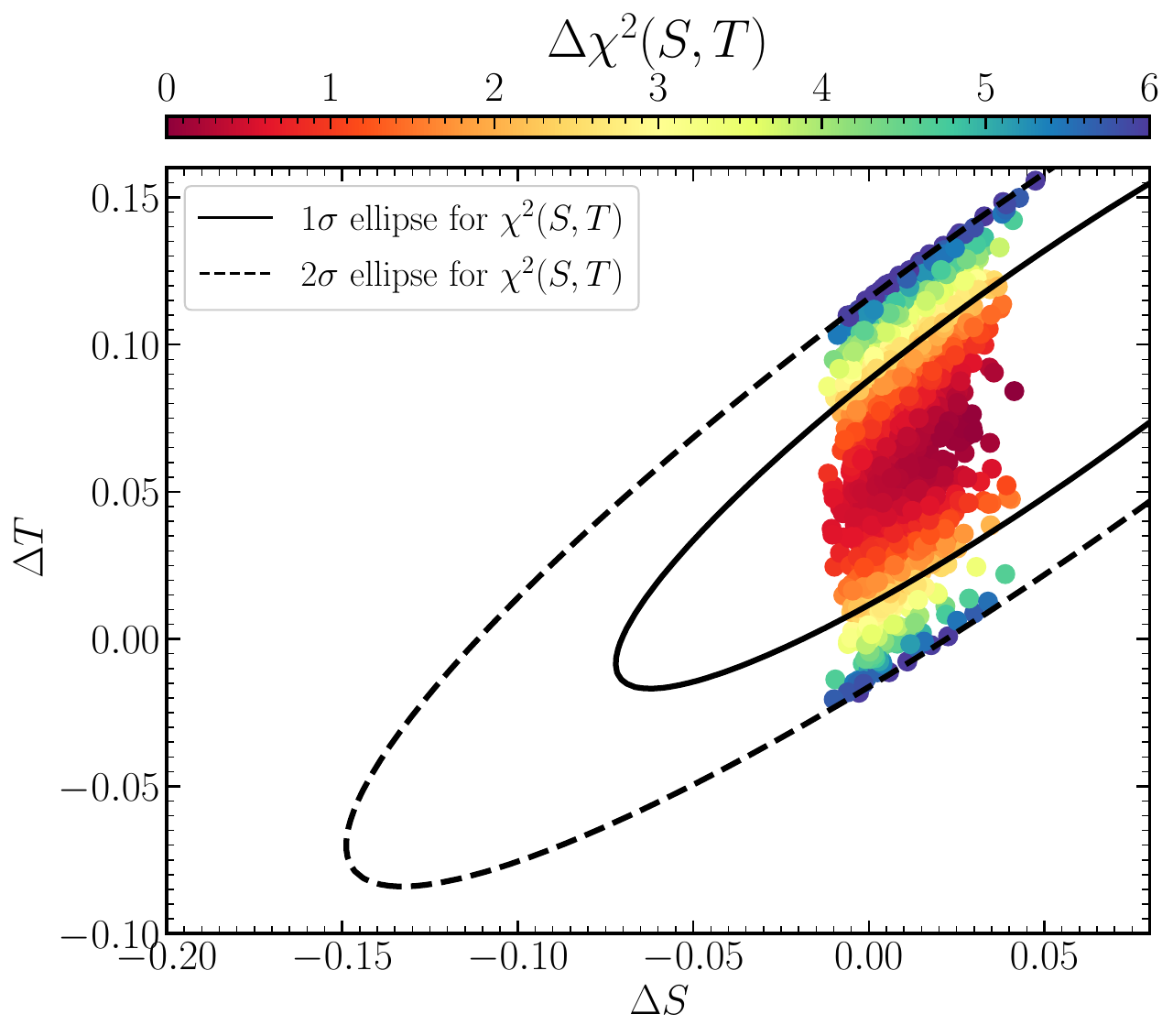}
	\caption{Allowed points by all constraints (\texttt{HiggsBounds}, \texttt{HiggsSignals}, \texttt{SuperIso} and theoretical ones) superimposed onto the fit limits in the
		$( S, ~ T)$ plane from EWPO data at 95\% CL (with a correlation of 92\%), with the colour
		code indicating $\Delta\chi^2(S,T)$. Results are presented for 
				the 2HDM-II+$(TB)$.}\label{fig6}
\end{figure}

Following the scan described in Tab. \ref{parameters}, we again present in Fig.~\ref{fig6} our 2HDM-II+$(TB)$ surviving points in the $(S,T)$ plane, wherein the colour illustrates the usual difference in  $\chi^2(S,~T)$ values. As previously, we will define the data set to be used for additional analysis as the one lying within the illustrated $2\sigma$ band. As evident, in this representation, the shape differs from the one presented in the previous representation, and this is again attributed to the cancellation of contributions from 2HDM-II and VLQ states to the oblique parameters $S$ and $T$.
	
 		\begin{figure}[H]
	\centering
	\includegraphics[height=12.1cm,width=16.5cm]{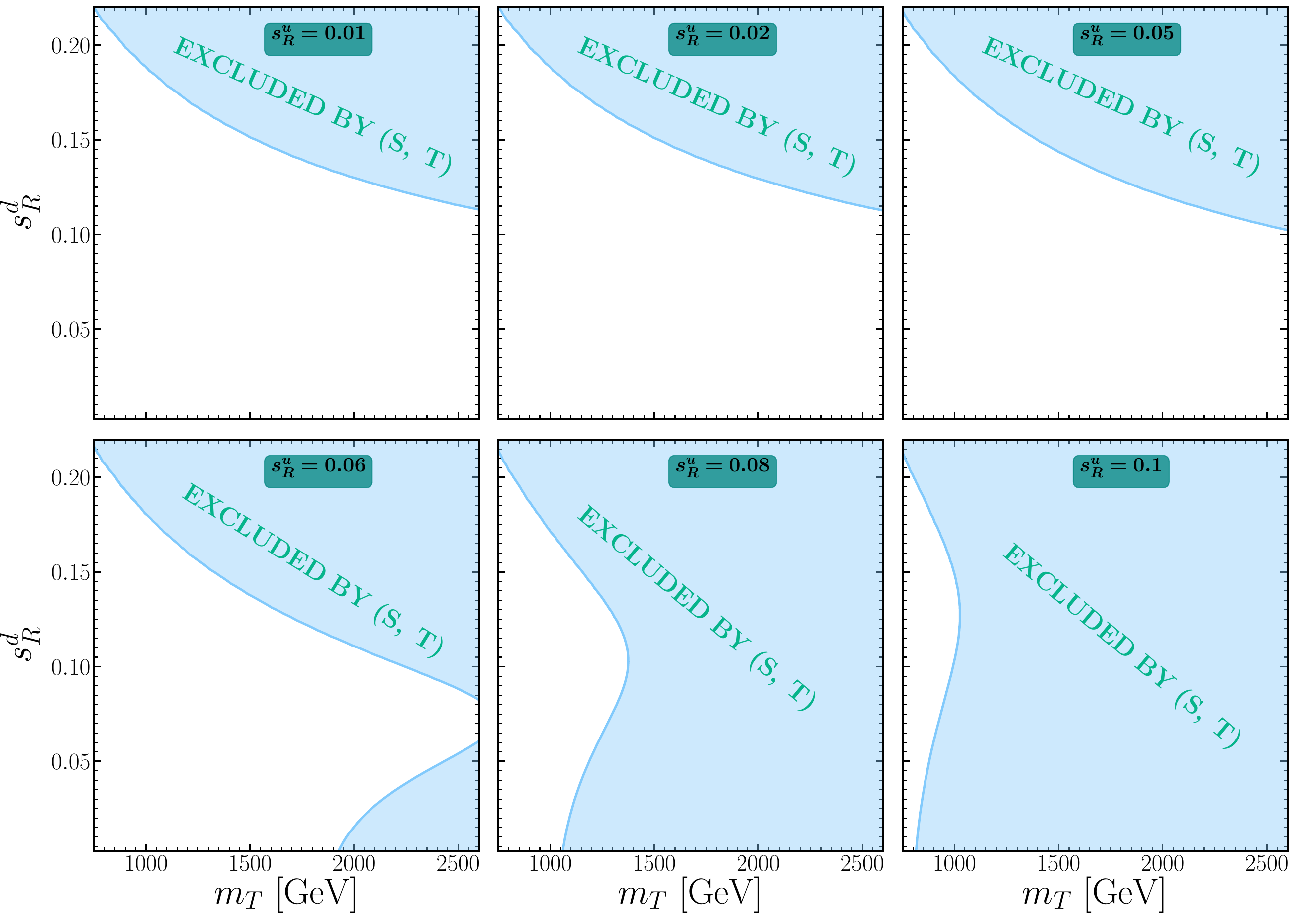}
	\caption{Exclusion regions by $(S,~T)$ in the $(m_T,~s^d_R)$ plane for various values of $s^u_R$, with fixed parameters: $m_{H}=532.80$, $m_A=524.26$ and $m_{H^\pm}=588.02$, with $\tan\beta=5.19$.}\label{fig9}
\end{figure}
Fig.~\ref{fig9} illustrates the exclusion region derived from the oblique parameters $S$ and $T$ at 2$\sigma$. It is evident from the plot that EWPOs impose a stringent limit on the mixing angles. Specifically, as the mixing angle $s^u_R$ increases, the exclusion region expands, covering the entire mass range of $m_T$. Similarly, for the mixing angle $s^d_R$, an increase in the VLQ mass $m_T$ results in a larger exclusion region.

\begin{figure}[H]
		\begin{minipage}{0.46\textwidth}
		\centering
		\includegraphics[height=9cm,width=8.3cm]{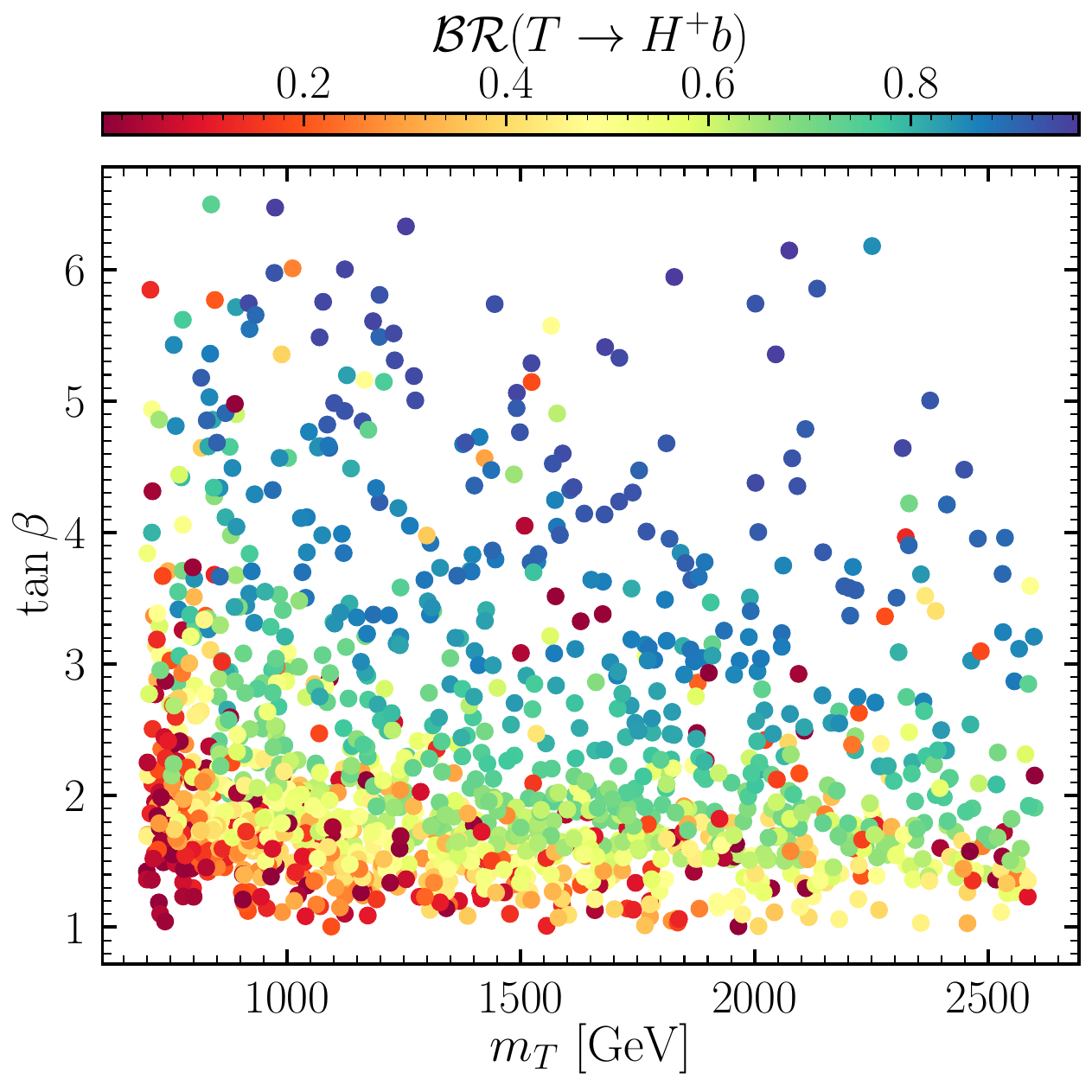}
	\end{minipage}\hspace{0.5cm}	
	\begin{minipage}{0.46\textwidth}
		\centering
		\includegraphics[height=9cm,width=8.3cm]{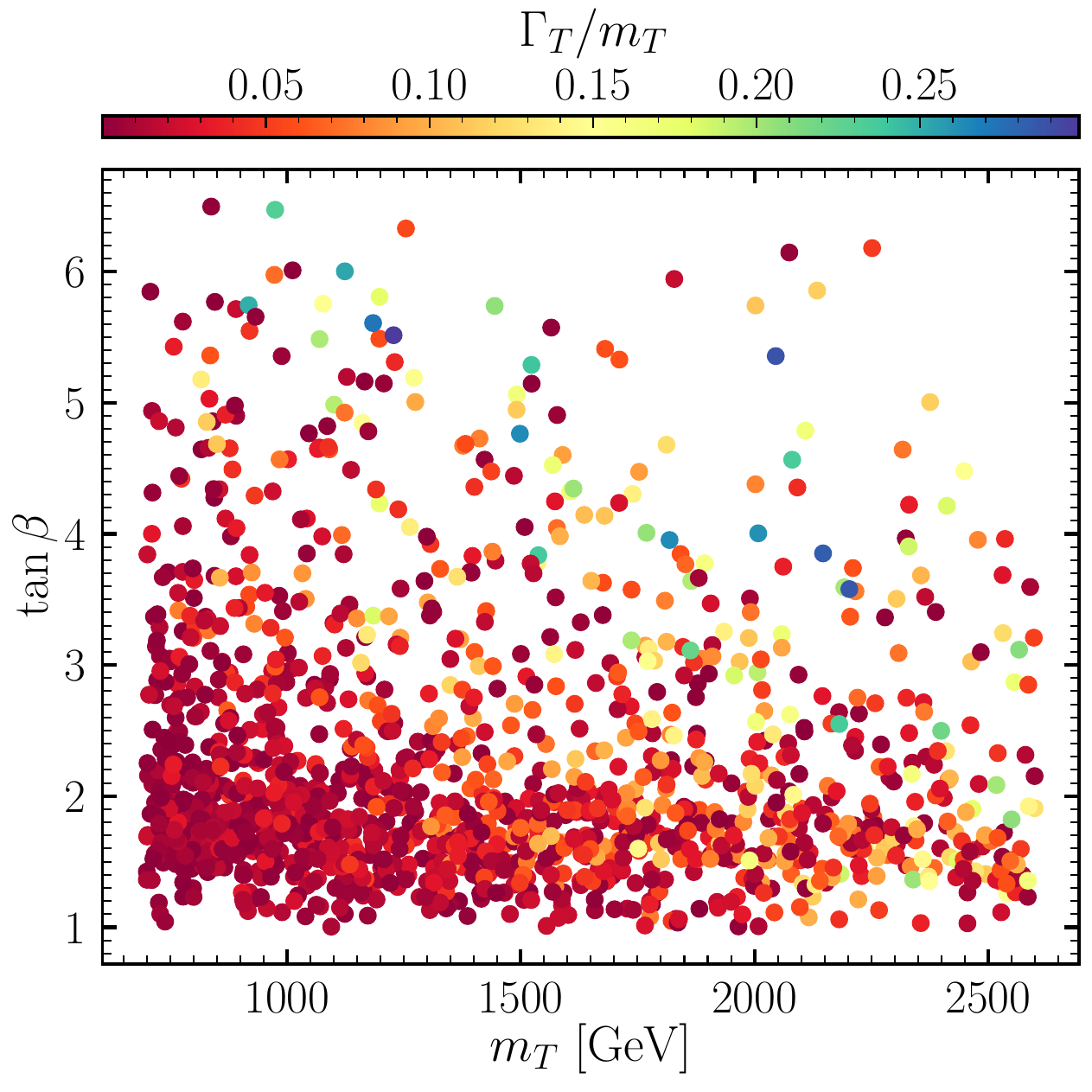}
	\end{minipage}
		
	\caption{$\mathcal{BR}(T\to H^+ b)$ (left) and  $\Gamma_T/m_T$ (right) plotted over the $(m_T,~\tan\beta)$ plane. Results are presented for 
				the 2HDM-II+$(TB)$.}\label{fig7}
\end{figure}

From Fig.~\ref{fig7} (left), in contrast to the previous VLQ realisation, it is clear that charged Higgs boson production from $T$ decays can reach more than 90\% in the corresponding ${\cal BR}$ for medium $\tan\beta$. In the right plot of the figure, we display $\Gamma_T/m_T$ as a function of $m_T$ and $\tan\beta$, where it can be seen that the total decay width of the $T$ state can reach 30\% of $m_T$. Hence, unlike the singlet case, here we are normally in presence of a rather sizeable $T\to H^\pm b$ rate, which is affected by phase space effects only for light $T$ states. Conversely, though, the $T$ state can be quite wide, thereby rendering attempts at reconstructing its mass from kinematic analysis potentially more difficult than in the singlet case.

\begin{figure}[H]
	\begin{minipage}{0.46\textwidth}
		\centering
		\includegraphics[height=9cm,width=8.3cm]{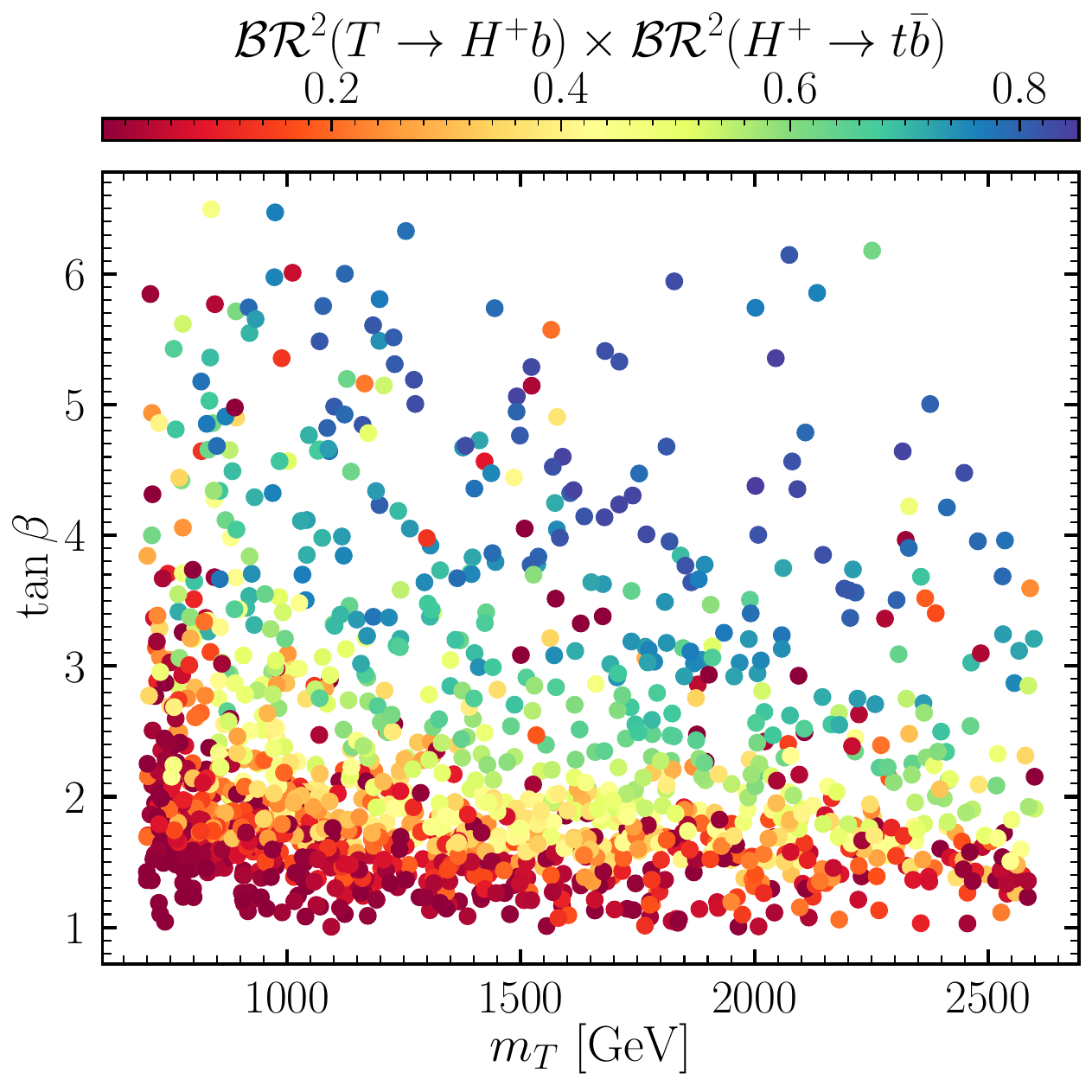}
	\end{minipage}\hspace{0.5cm}	
	\begin{minipage}{0.46\textwidth}
		\centering
		\includegraphics[height=9cm,width=8.3cm]{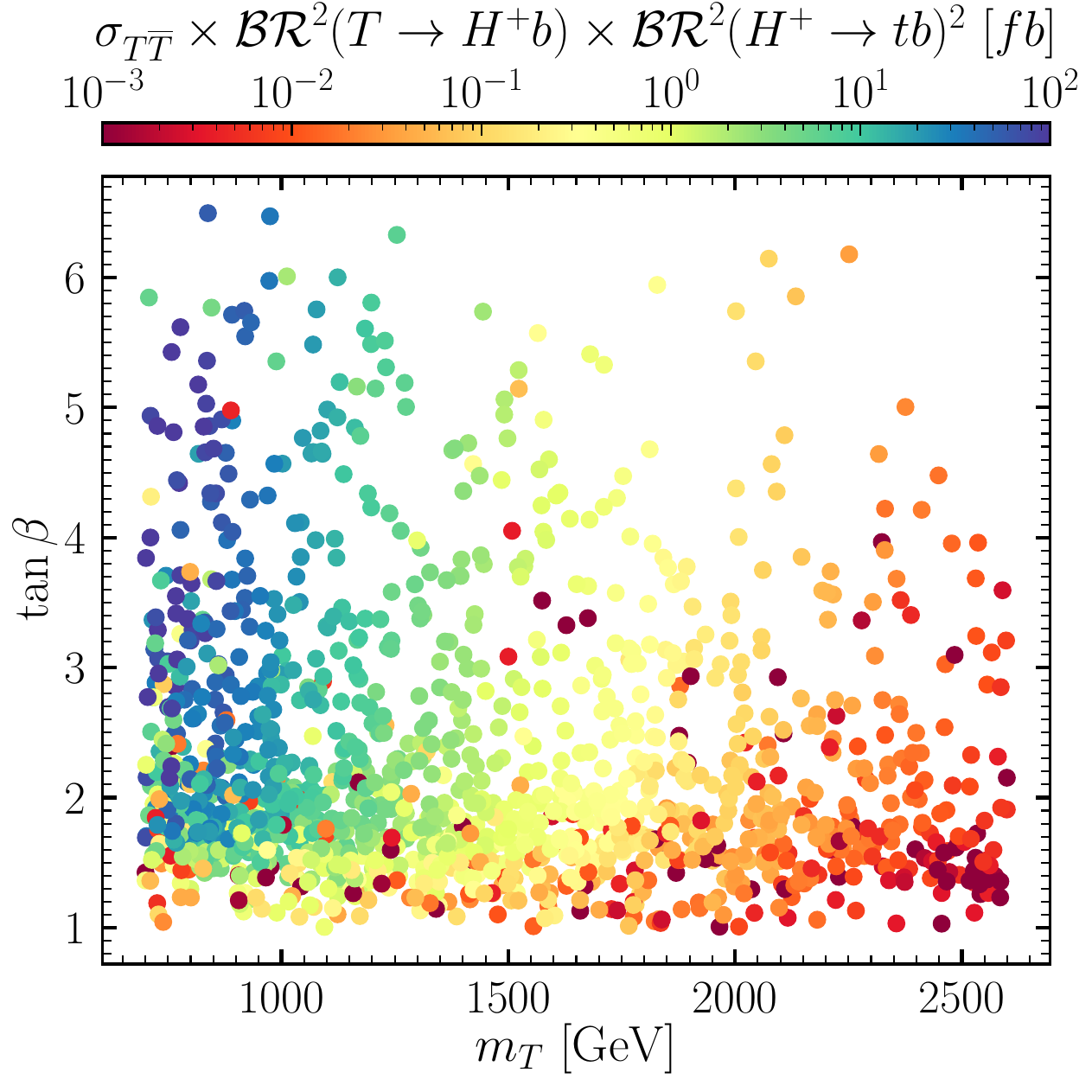}
	\end{minipage}
	
	\caption{$\mathcal{BR}^2(T\to H^+ b)\times\mathcal{BR}^2(H^+\to tb)$ (left) and  $\sigma_{T\overline{T}}\times\mathcal{BR}^2(T\to H^+ b)\times\mathcal{BR}^2(H^+\to tb)$ (right) plotted over the $(m_T,~\tan\beta)$ plane. Results are presented for 
				the 2HDM-II+$(TB)$.}\label{fig8}
\end{figure}

In Fig.~\ref{fig8}, we present in the ($m_T, \tan\beta$) plane the aforementioned distribution of points mapped against $\mathcal{BR}^2(T\to H^+ b)\times\mathcal{BR}^2(H^+\to tb)$ (left panel) and $\sigma_{T\overline{T}}\times\mathcal{BR}^2(T\to H^+ b)\times\mathcal{BR}^2(H^+\to tb)$ (right panel). One can see from these plots that the signal $2t4b$ could reach values up to 100 fb for medium $\tan\beta$ and for $m_T\le1000$ GeV. Therefore, owing to the enhanced coupling $TH^+b$ in this scenario, a much larger cross section is observed for this signature compared to the 2HDM-II+($T$) case.

	\subsection{BPs}
Before concluding, in order to encourage experimental analyses of these scenarios at the LHC, we propose three BPs in Tab.~\ref{BP_T} for the 2HDM-II+$(TB)$ representation, which exhibits a more substantial cross-section compared to the 2HDM-II+$(T)$ one. The chosen BPs feature $T$ masses within the range of approximately 800 GeV to 1300 GeV. Furthermore, the $H^\pm$ masses are set to be heavy, enabling attempts to extract charged Higgs boson signatures in its $tb$ decay, presenting a significant signal in the form of $2t4b$ final states originating from the pair production of the new VLQ $T$, followed bt $T\to H^+b$ and $H^+\to  tb$ decays.

		\section{Conclusions}

This paper goes into great detail about how charged Higgs bosons are made with a new top partner $T$ in two different forms: 2HDM-II+$(T)$ (singlet) and 2HDM-II+$(TB)$ (doublet). Our findings reveal a compelling capability of VLQs to alleviate the stringent limit imposed by $B\to X_s\gamma$ on the charged Higgs mass ($m_{H^\pm}$). Specifically, in the singlet scenario, the mass limit can be reduced to around 500 GeV, while in the doublet case, it can decrease to approximately 200 GeV,particularly when the mixing angles are large. However, the inclusion of constraints from the oblique parameters $S$ and $T$ implies additional limitations, allowing only small VLQ mixing angles, in turn somewhat restraining the reduction in the aforementioned limit on $m_{H^\pm}$. Consequently, in the 2HDM-II+$(T)$ scenario, the mass limit decreases to about 567 GeV, and in the 2HDM-II+$(TB)$ scenario, it reduces to approximately 360 GeV, compared to the typical 580 GeV in the 2HDM-II.

Furthermore, we observed a remarkable distinction between the 2HDM-II+$(TB)$ and 2HDM-II+$(T)$ representations. The former exhibits a substantial production rate of charged Higgs bosons, approaching nearly 100\%, while the latter achieves only 25\%, in terms of the ${\cal BR}(T\to H^+b)$. Our study thus further delved into the pair production of $T$ ($pp\to T\bar{T}$), followed by the decays $T\to H^\pm b$ of both VLQs. In both scenarios, the charged Higgs boson subsequently undergoes decay into $tb$, yielding a distinctive final state characterised by two top and four bottom quarks ($2t4b$). Notably, our results indicate that this signal can reach 100 fb in the 2HDM-II+$(TB)$ whereas for the 2HDM-II+$(T)$ case, the corresponding rates are significantly smaller, so as to be of little relevance for the forthcoming LHC runs. Hence, we finally produced three BPs in the 2HDM-II+$(TB)$ amenable to experimental investigation.

While a comprehensive analysis of $H^\pm$ decays has not been performed in this study, we are confident that the $H^\pm\to tb$ channel ($2t4b$ final state) from $T\bar T$ production and decay is a promising avenue for further exploration. This endeavour, most likely during Run 3 and certainly at the HL-LHC, awaits a future publication where a detailed analysis of $H^\pm$ decay channels will be performed.

\color{black}
\noindent{\bf Acknowledgments}

\noindent
 SM is supported in part through the NExT Institute and the STFC Consolidated Grant No. ST/L000296/1. We thank Rikard Enberg for many fruitful discussions.

\begin{table}[H]
	\begin{center}
		\setlength{\tabcolsep}{24pt}
		\renewcommand{\arraystretch}{0.95}
		\begin{adjustbox}{max width=\textwidth}		
			\begin{tabular}{lccc}
				\hline\hline
				\multirow{2}{*}{Parameters}& \multicolumn{3}{c}{2HDM-II+$(TB)$} \\
				&       BP$_1$ &       BP$_2$&       BP$_3$ \\
				\hline\hline

				$m_h$   &  125.00 &  125.00 &   125.00  \\
				$m_H$   &  425.91 &  572.13 &  532.80  \\
				$m_A$   &  414.07&  568.50 &  524.26  \\
				$m_{H^\pm}$   &   448.66 & 546.59 &  588.02  \\
				
				$\tan\beta$ &   4.85&  6.00 &    5.19  \\
				
				$m_T$      & 828.66& 1123.98 & 1271.68  \\
				$m_B$      &  841.84& 1137.92 & 1282.16 \\
				$s_u^L$    &    $-0.0201$& $-0.0089$ &    0.0018  \\
				$s_d^L$    &    $-0.0011$& $-0.0007$ &    0.0005  \\
				$s^u_R$    &    $-0.0964$& $-0.0579$ &    0.0129  \\
				$s^d_R$     &    $-0.1992$& $-0.1660$ &    0.1283  \\

				\hline\hline
				\multicolumn{4}{c}{$\mathcal{BR}(H^\pm\to XY)$ in \%} \\\hline\hline
				${\cal BR}(H^+\to t\bar{b})$ &   96.97&  93.83 & 96.17 \\
				\hline\hline
				\multicolumn{4}{c}{$\mathcal{BR}(T\to XY)$ in \%} \\\hline\hline
				${\cal BR}(T\to W^+b)$ & 7.74 & 4.53 &  5.66  \\
				${\cal BR}(T\to Zt)$   & 0.79& 0.26 &  0.03  \\
				${\cal BR}(T\to ht)$  &  1.04& 0.30 &  0.03  \\
				${\cal BR}(T\to H^+b)$ &90.43& 94.92 & 94.28  \\   
				\hline\hline
				
				\multicolumn{4}{c}{$\Gamma$ in $\mathrm{GeV}$} \\\hline\hline
				$\Gamma(T)$  & 95.64 & 283.32 & 196.11   \\ 
				\hline\hline
				\multicolumn{4}{c}{$\sigma$ [fb]} \\\hline\hline

				$\sigma_{T\overline{T}}\times{\cal BR}^2(T\to H^+b)\times{\cal BR}^2(H^+\to t\bar{b})$ & 99.86 & 14.09 & 6.22  \\ 

				\hline\hline%
			\end{tabular}
		\end{adjustbox}
	\end{center}
	\caption{The full description of our BPs. Masses are in GeV.}\label{BP_T}
\end{table}

\clearpage
\appendix
\section{Charged Higgs Boson Couplings}\label{App_charged}
In this section, we provide further elaboration on the charged Higgs couplings within the 2HDM-II+$(T)$ and 2HDM-II+$(TB)$ frameworks.
\subsection{2HDM-II+${(T)}$}
The Yukawa Lagrangian of the 2HDM-II+$(T)$ can be expressed as\footnote{Note that in the 2HDM-II: $\Phi_1=\begin{pmatrix}
	H^-_d\\-H^0_d
	\end{pmatrix}$ couples to down quarks and $\Phi_2=\begin{pmatrix}
	H^0_u\\-H^+_u
	\end{pmatrix}$ couples to up quarks.}:

\begin{equation}
{    \mathcal{L}_Y=- \yu\Ql\tilde{\Phi}_2\ur+\yd\Ql\Phi_1\dr -y_{i4}^{u} \Ql \tilde{\Phi}_2u_{R4}^0 + h.c.}\label{eq1}
\end{equation}

where ${\Phi}_{1,2}$ are the Higgs doublets, $\tilde{\Phi}_i \equiv i\sigma_2\Phi^*_i$, $\overline{Q}^0_{Li}= \begin{pmatrix}
\uli &\dli
\end{pmatrix}$ is the weak isospin quark doublet, and $\ur$, $u_{R4}^0$ and $\dr$ are weak isospin quark singlets.

\noindent Assuming that the new VLQs predominantly mix with the third generation ($i,j = 3$), we can express this as:
\begin{align}
\mathcal{L}_Y=&- \yt\Qlm\hum\urt\nonumber\\&-{y_{34}^{u}} \Qlm \hum\T \nonumber\\&+\yb\Qlm\hdm\drb+h.c.
\end{align}	
Let's focus on the charged Higgs boson $H^\pm$:
\begin{eqnarray}
\mathcal{L}_Y&\supset & \yt\Qlm\begin{pmatrix}
0\\\cos\beta H^\pm
\end{pmatrix}\urt\nonumber\\&+&{y_{34}^{u}} \Qlm  \begin{pmatrix}
0\\\cos\beta H^\pm
\end{pmatrix}\T\nonumber\\&+&\yb\Qlm\begin{pmatrix}
\sin\beta H^\pm\\0
\end{pmatrix}\drb\nonumber\\
&\supset&\yt \dlb\urt\cos\beta H^\pm+{y_{34}^{u}}\dlb\T\cos\beta H^\pm +\yb \ult\drb\sin\beta H^\pm
\end{eqnarray}

\begin{equation}
\mathcal{L}_Y= \dLm\underbrace{\yuu}_{(*)}\uRm\cos\beta H^\pm+y_{33}^d\uLm \dRm\sin\beta H^\pm 
\label{Lag}\end{equation}
Given our assumption that the new VLQs primarily mix with the third generation, we can write:
\begin{equation}
\left(\! \begin{array}{c} u_{L3,R3} \\ u_{L4,R4} \end{array} \!\right) =
U_{L,R}^u \left(\! \begin{array}{c} u^0_{L3,R3} \\ u^0_{L4,R4} \end{array} \!\right)
= \left(\! \begin{array}{cc} c_{L,R}^u & -s_{L,R}^u e^{i \phi_u} \\ s_{L,R}^u e^{-i \phi_u} & c_{L,R}^u \end{array}
\!\right)
\left(\! \begin{array}{c} u^0_{L3,R3} \\ u^0_{L4,R4} \end{array} \!\right) \,.
\label{ec:mixu1}
\end{equation}
Now, let's express $(*)$ in the form of the mass matrix $\mathcal{M}^u$:

\begin{eqnarray}
\yuu&=&\frac{\sqrt{2}}{v}\underbrace{\begin{pmatrix}
	1&&0\\0&&0
	\end{pmatrix}}_{Y^0}\underbrace{\begin{pmatrix}
	\yt\frac{v}{\sqrt{2}}&\ytt\frac{v}{\sqrt{2}}\\0&M^0 
	\end{pmatrix}}_{{\mathcal{M}^u} }\label{eq6}
\end{eqnarray}

\noindent Consequently, Eq. (\ref{Lag}) can be formulated as:

\begin{eqnarray}
\mathcal{L}_Y&=& \frac{\sqrt{2}}{v}\Biggl[\dLmr U^d_L Y^0 {U_L^u}^\dagger \underbrace{U_L^u \mathcal{M}^u {U_R^u}^\dagger}_{\mathcal{M}_{diag}^u} U_R^u {U_R^u}^\dagger \uRmr \cos\beta\nonumber\\&&+m_b\left(\clu \bar{u}_{L3}d_{R3}+\slu \bar{u}_{L4}d_{R3}\right)\sin\beta\Biggr] H^\pm\nonumber\\\label{eq8}
&=& 
\frac{\sqrt{2}}{v}\Biggl[ \dLmr\begin{pmatrix}
m_tc^u_L&m_T s^u_L\\0&0
\end{pmatrix}\uRmr\cos\beta\nonumber\\&&+m_b\left(\clu \bar{u}_{L3}d_{R3}+\slu \bar{u}_{L4}d_{R3}\right)\sin\beta\Biggr] H^\pm
\end{eqnarray}
Subsequently, we obtain:
\begin{eqnarray}
H^\pm tb&=&-\frac{g}{\sqrt{2}M_W\sin\beta}m_tc^u\cos\beta-\frac{g}{\sqrt{2}M_W\cos\beta}m_bc^u_L\sin\beta\nonumber\\&=&-\frac{gm_t}{\sqrt{2}M_W}\Biggl[  c^u_L\cot\beta+ \frac{m_b}{m_t} c^u_L \tan\beta \Biggr]\\
H^\pm Tb&=&-\frac{g}{\sqrt{2}M_W\sin\beta}m_Ts^u_L\cos\beta-\frac{g}{\sqrt{2}M_W\cos\beta}m_bs^u_L\sin\beta\nonumber\\&=&-\frac{gm_T}{\sqrt{2}M_W}\Biggl[  s^u_L\cot\beta+ \frac{m_b}{m_T} s^u_L \tan\beta \Biggr]
\end{eqnarray}

\noindent Hence:

\begin{eqnarray}
\begin{array}{cccc}
Z_L^{tb}= c^u_L &&& Z_R^{tb}=\frac{m_b}{m_t} c^u_L\\
Z_L^{Tb}= s^u_L &&& Z_R^{Tb}=\frac{m_b}{m_T} s^u_L\\
\end{array} 
\end{eqnarray}
\begin{figure}[H]
	\centering
	\includegraphics[height=7.5cm,width=16cm]{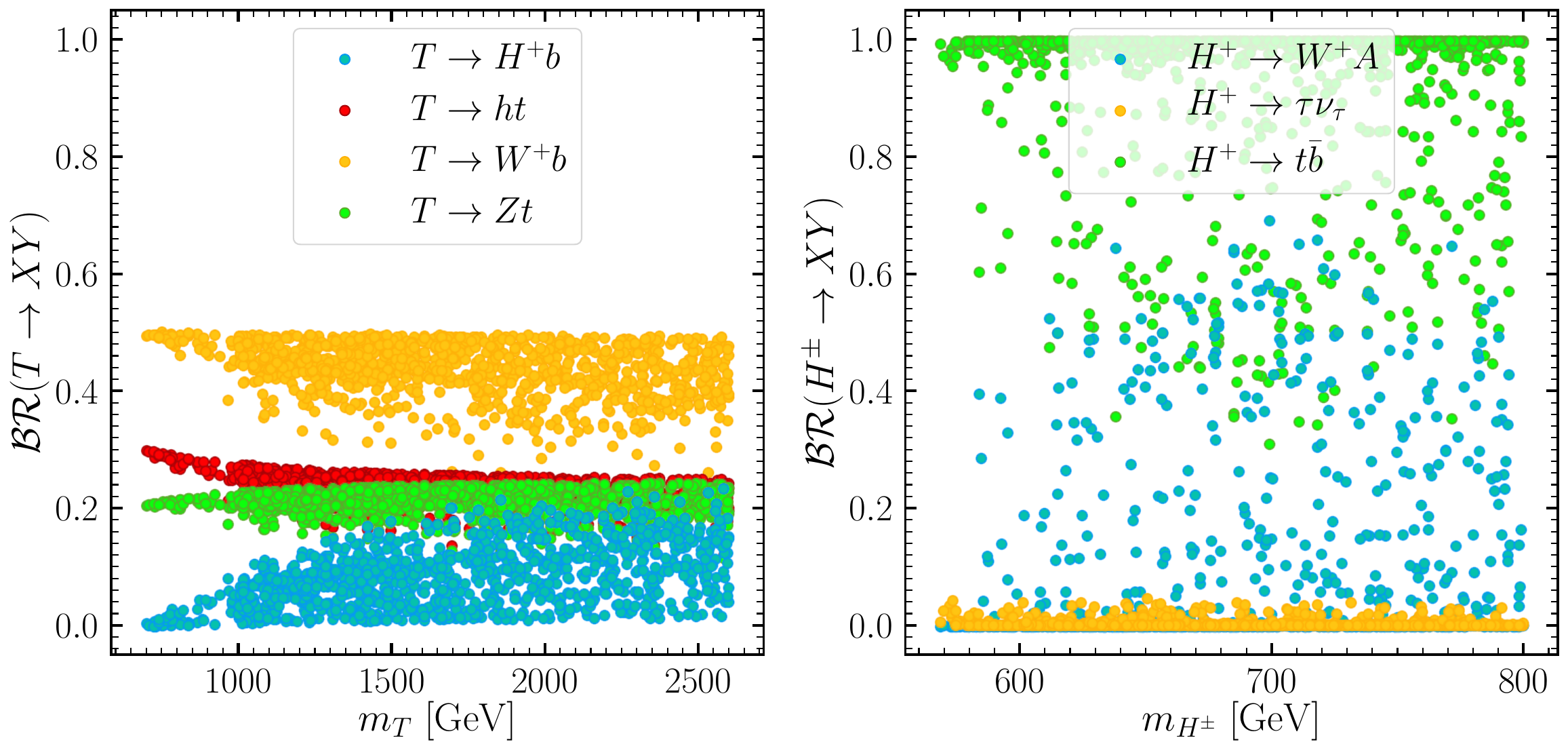}

	\caption{Left:$\mathcal{BR}(T\to XY)$ as a function of $m_T$, with   XY= $H^+b$ blue, $ht$ red, $W^+b$ yellow and $Zt$ green. Right: $\mathcal{BR}(H^\pm\to XY)$ as a function of $m_{H^\pm}$, with   XY= $W^+A$ blue, $\tau\nu$ yellow, and $tb$ green. For 2HDM-II+$(T)$. }\label{fig11}
\end{figure}
In Fig.~\ref{fig11}, the {\cal BR}s of $T$ and $H^\pm$ in the 2HDM-II+$(T)$ scenario are illustrated as functions of their respective masses. The left panel indicates that the $\mathcal{BR}(T\to H^+ b)$ can reach a maximum of 25\% for $m_T > 2000$ GeV. This limitation is due to the coupling depicted earlier, which is proportional to the mixing angle $s^u_L$, a parameter constrained to be very small by the oblique parameters $S$ and $T$. Shifting to the right panel, where clearly we can see that  the charged Higgs boson predominantly decays through the fermionic channel $H^+\to t\bar{b}$ achieving approximately 100\% across the entire range of charged Higgs masses.
\subsection{2HDM-II+${(TB)}$}
The Yukawa Lagrangian for the doublet $(TB)$ case can be expressed as:
\begin{equation}
{	\mathcal{L}_Y=- \yu\Qlj\tilde{\Phi}_2\ur+\yd\Qlj\Phi_1\dr -y_{4j}^{u} \Qlj \tilde{\Phi}_2u_{R4}^0 +y_{4j}^{d} \Qlj {\Phi}_1d_{R4}^0+h.c.}\label{eq2}
\end{equation}
Following the same steps as in the previous scenario, we can express the charged Higgs Yukawa Lagrangian as:
\begin{eqnarray}
\mathcal{L}_{H^\pm}=\uLm\underbrace{\begin{pmatrix}
y_{33}^d&0\\y_{43}^d&0
\end{pmatrix}}_{*}\dRm\sin\beta H^\pm+\dLm\underbrace{\begin{pmatrix}
y_{33}^u&0\\y_{43}^u&0
\end{pmatrix}}_{**}\begin{pmatrix}
u_{R3}^0\\u_{R4}^0
\end{pmatrix}\cos\beta H^\pm\label{yukTB}
\end{eqnarray}
Now, let's express $(*)$ and $(**)$ in the form of the mass matrices $\mathcal{M}^u$ and $\mathcal{M}^d$:
\begin{eqnarray}
\begin{pmatrix}
y_{33}^d&0\\y_{43}^d&0
\end{pmatrix}&=&\frac{\sqrt{2}}{v}\underbrace{\begin{pmatrix}
y_{33}^d\frac{v}{\sqrt{2}}&0\\y_{43}^d\frac{v}{\sqrt{2}}&M^0
\end{pmatrix}}_{\mathcal{M}^d}\underbrace{\begin{pmatrix}
1&0\\0&0
\end{pmatrix}}_{Y^0}
\end{eqnarray}

\begin{eqnarray}
\begin{pmatrix}
y_{33}^u&0\\y_{43}^u&0
\end{pmatrix}&=&\frac{\sqrt{2}}{v}\underbrace{\begin{pmatrix}
	y_{33}^u\frac{v}{\sqrt{2}}&0\\y_{43}^u\frac{v}{\sqrt{2}}&M^0
	\end{pmatrix}}_{\mathcal{M}^u}\underbrace{\begin{pmatrix}
	1&0\\0&0
	\end{pmatrix}}_{Y^0}
\end{eqnarray}
Then Eq. (\ref{yukTB}) can be written as:
\begin{eqnarray}
\mathcal{L}_{H^\pm}&=&\frac{\sqrt{2}}{v}\Biggl[\uLm\mathcal{M}^dY^0\dRm\sin\beta +\dLm\mathcal{M}^uY^0\begin{pmatrix}
u_{R3}^0\\u_{R4}^0
\end{pmatrix}\cos\beta \Biggr]H^\pm\nonumber\\&=&\frac{\sqrt{2}}{v}\Biggl[\uLmr U^u_L{U^d_L}^\dagger\underbrace{ U^u_L\mathcal{M}^u {U^u_R}^\dagger}_{\mathcal{M}^d_{diag}} U_R^dY^0 {U_R^d}^\dagger\dRmr\sin\beta \nonumber\\&&+\dLmr U^d_L{U^u_L}^\dagger\underbrace{ U^u_L\mathcal{M}^u {U^u_R}^\dagger}_{\mathcal{M}^u_{diag}} U_R^uY^0{U_R^u}^\dagger\begin{pmatrix}
u_{R3}\\u_{R4}
\end{pmatrix}\cos\beta\Biggr] H^\pm\nonumber\\&=&\frac{\sqrt{2}}{v}\Biggl[\uLmr \underbrace{\begin{pmatrix}
\clu\cld+\slu\sld&\clu\sld-\slu\cld\\\slu\cld-\clu\sld&\slu\sld+\clu\cld
\end{pmatrix} \begin{pmatrix}
{\crd}^2m_b&\srd\crd m_b\\ \srd\crd m_B&{\srd}^2m_B
\end{pmatrix}}_{\mathcal{M}_1}\dRmr\sin\beta \nonumber\\&&+\dLmr \underbrace{\begin{pmatrix}
\cld\clu+\sld\slu&\cld\slu-\sld\clu\\\sld\clu-\cld\slu&\sld\slu+\cld\clu
\end{pmatrix} \begin{pmatrix}
{\cru}^2m_t&\sru\cru m_t\\ \sru\cru m_T&{\sru}^2m_T
\end{pmatrix}}_{\mathcal{M}_2}\begin{pmatrix}
u_{R3}\\u_{R4}
\end{pmatrix}\cos\beta\Biggr] H^\pm\nonumber\\
\end{eqnarray}
For the coupling of $H^+tb$:

\begin{eqnarray}
H^\pm tb&=& -\frac{g}{\sqrt{2}M_W\cos\beta}\sin\beta\mathcal{M}_1[1,1]-\frac{g}{\sqrt{2}M_W\sin\beta}\cos\beta\mathcal{M}_2[1,1]\nonumber\\&=&-\frac{g\tan\beta}{\sqrt{2}M_W}\Bigl[m_b{\crd}^2\left(\clu\cld+\slu\sld\right)+m_B\srd\crd\left(\clu\sld-\slu\cld\right)\Bigr]\nonumber\\&&-\frac{g\cot\beta}{\sqrt{2}M_W}\Bigl[m_t{\cru}^2\left(\cld\clu+\sld\slu\right)+m_T\sru\cru\left(\cld\slu-\sld\clu\right)\Bigr]
\end{eqnarray}
we use :
\begin{eqnarray}
c^{u,d}_R=\frac{m_q}{m_Q}\frac{c^{u,d}_L}{s^{u,d}_L}s^{u,d}_R\label{eq28}
\end{eqnarray}
we get:

\begin{eqnarray}
H^\pm tb&=&-\frac{g m_t}{\sqrt{2}M_W}\Bigl[{\cld\clu+\frac{\sld}{\slu}\left({\slu}^2-{\sru}^2\right)}\cot\beta+{\frac{m_b}{m_t}\Big[\clu\cld+\frac{\slu}{\sld}\left({\sld}^2-{\srd}^2\right)\Big]}\tan\beta\Bigr]
\end{eqnarray}
then we get:
\begin{eqnarray}
\begin{array}{cccc}
Z_L^{tb}= {\cld\clu+\frac{\sld}{\slu}\left({\slu}^2-{\sru}^2\right)} &&& Z_R^{tb}={\frac{m_b}{m_t}\Big[\clu\cld+\frac{\slu}{\sld}\left({\sld}^2-{\srd}^2\right)\Big]}\\

\end{array} 
\end{eqnarray}
For the coupling of $H^\pm T b$ :
\begin{eqnarray}
H^\pm Tb&=& -\frac{g}{\sqrt{2}M_W\cos\beta}\sin\beta\mathcal{M}_1[2,1]-\frac{g}{\sqrt{2}M_W\sin\beta}\cos\beta\mathcal{M}_2[1,2]\nonumber\\&=&-\frac{g\tan\beta}{\sqrt{2}M_W}\Bigl[m_b{\crd}^2\left(\slu\cld-\clu\sld\right)+m_B\srd\crd\left(\slu\sld-\clu\cld\right)\Bigr]\nonumber\\&&-\frac{g\cot\beta}{\sqrt{2}M_W}\Bigl[m_t{\sru\cru}\left(\cld\clu+\sld\slu\right)+m_T{\sru}^2\left(\cld\slu-\sld\clu\right)\Bigr]
\end{eqnarray}
we use Eq.  (\ref{eq28}) we get:

\begin{eqnarray}
H^\pm Tb&=&-\frac{g m_t}{\sqrt{2}M_W}\Bigl[{\cld\slu+\frac{\sld}{\clu}\left({\slu}^2-{\sru}^2\right)}\cot\beta+{\frac{m_b}{m_T}\Big[\cld\slu+\frac{\clu}{\sld}\left({\srd}^2-{\sld}^2\right)\Big]}\tan\beta\Bigr]\nonumber\\
\end{eqnarray}
Finally we get:
\begin{eqnarray}
\begin{array}{cccc}
Z_L^{Tb}= {\cld\slu+\frac{\sld}{\clu}\left({\slu}^2-{\sru}^2\right)} &&& Z_R^{Tb}={\frac{m_b}{m_T}\Big[\cld\slu+\frac{\clu}{\sld}\left({\srd}^2-{\sld}^2\right)\Big]}\\
\end{array} 
\end{eqnarray}
\begin{figure}[H]
	\centering
	\includegraphics[height=7.5cm,width=16cm]{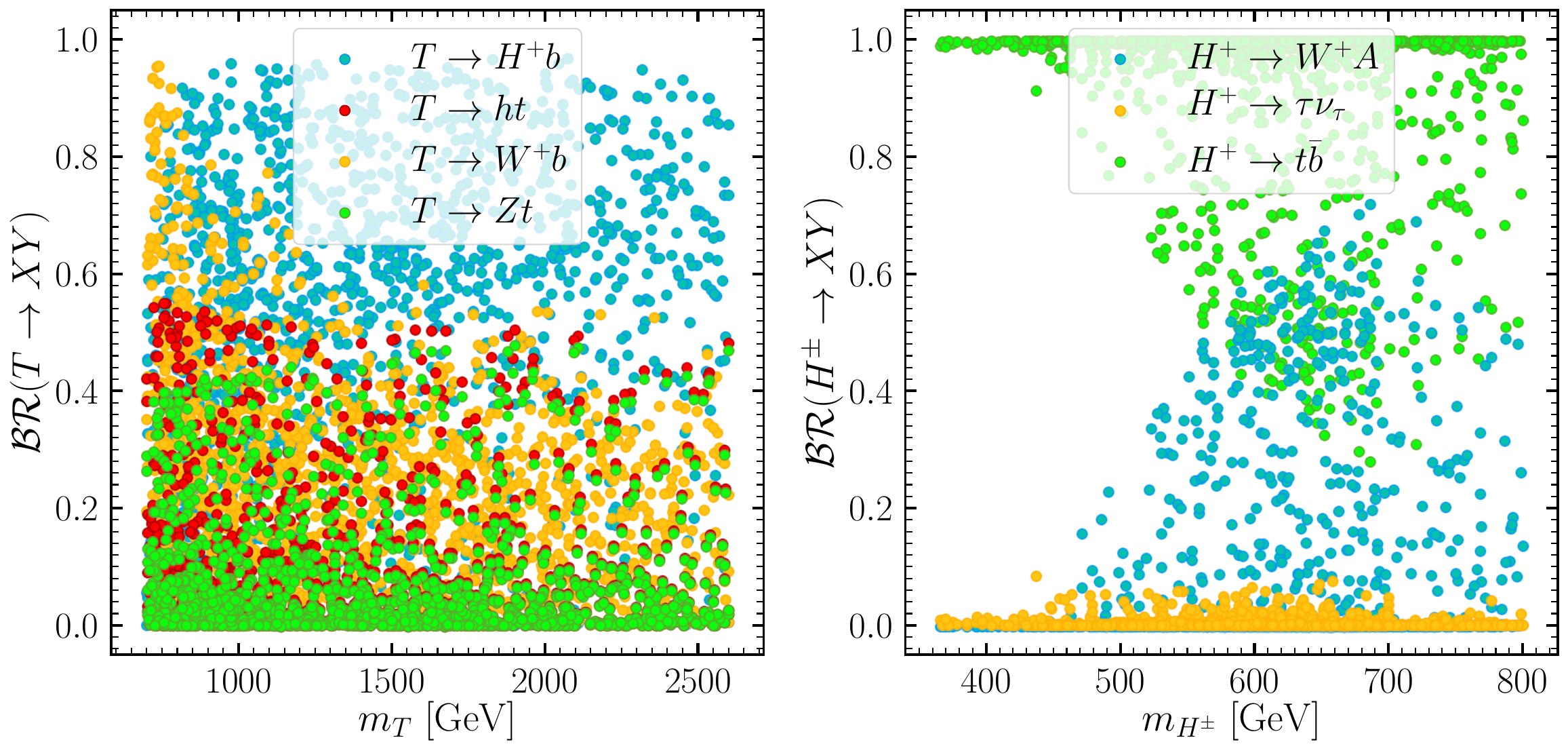}

	\caption{The same as Fig.~\ref{fig11} but for 2HDM-II+$(TB)$. }\label{fig12}
\end{figure}

Similar to Fig.~\ref{fig11}, we illustrate in Fig.~\ref{fig12} the ${\cal BR}$s of $T$ and $H^\pm$ as functions of their respective masses, but within the context of the 2HDM-II+$(TB)$ scenario. Beginning with the left panel, in contrast to the previous scenario (2HDM-II+$(T)$), the production of charged Higgs from the new top quark $T$ dominates for $m_T > 1000$ GeV, reaching nearly 100\%. Shifting to the right panel, and as anticipated for heavy charged Higgs masses, the primary decay channel remains the fermionic one $H^+\to t\bar{b}$, achieving approximately 100\% ${\cal BR}$ across the entire range of charged Higgs masses.
\section{LHC Limits on VLQs}
\begin{figure}[H]
	\centering
	\includegraphics[height=6.75cm,width=7.75cm]{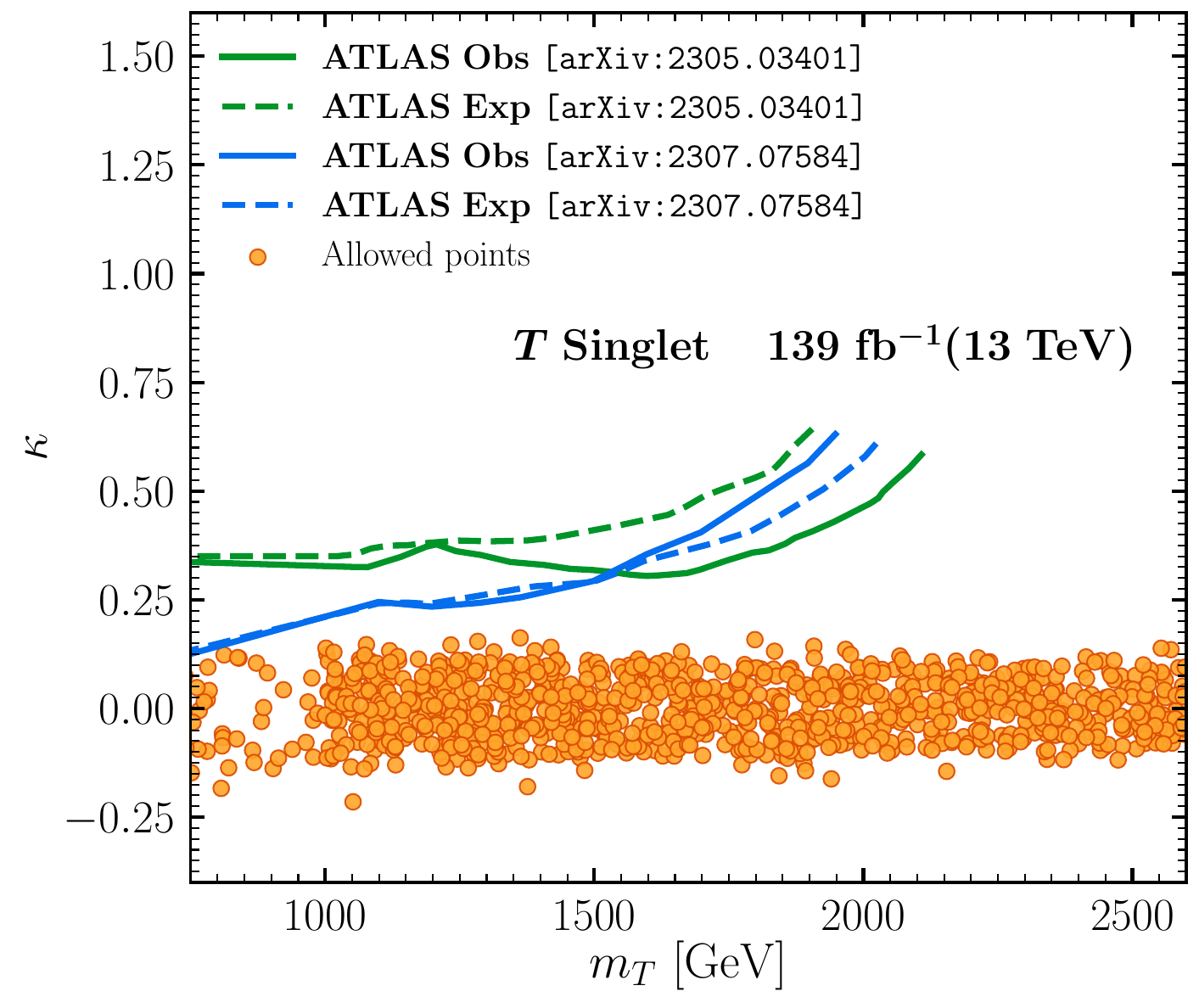}
	\includegraphics[height=6.75cm,width=7.75cm]{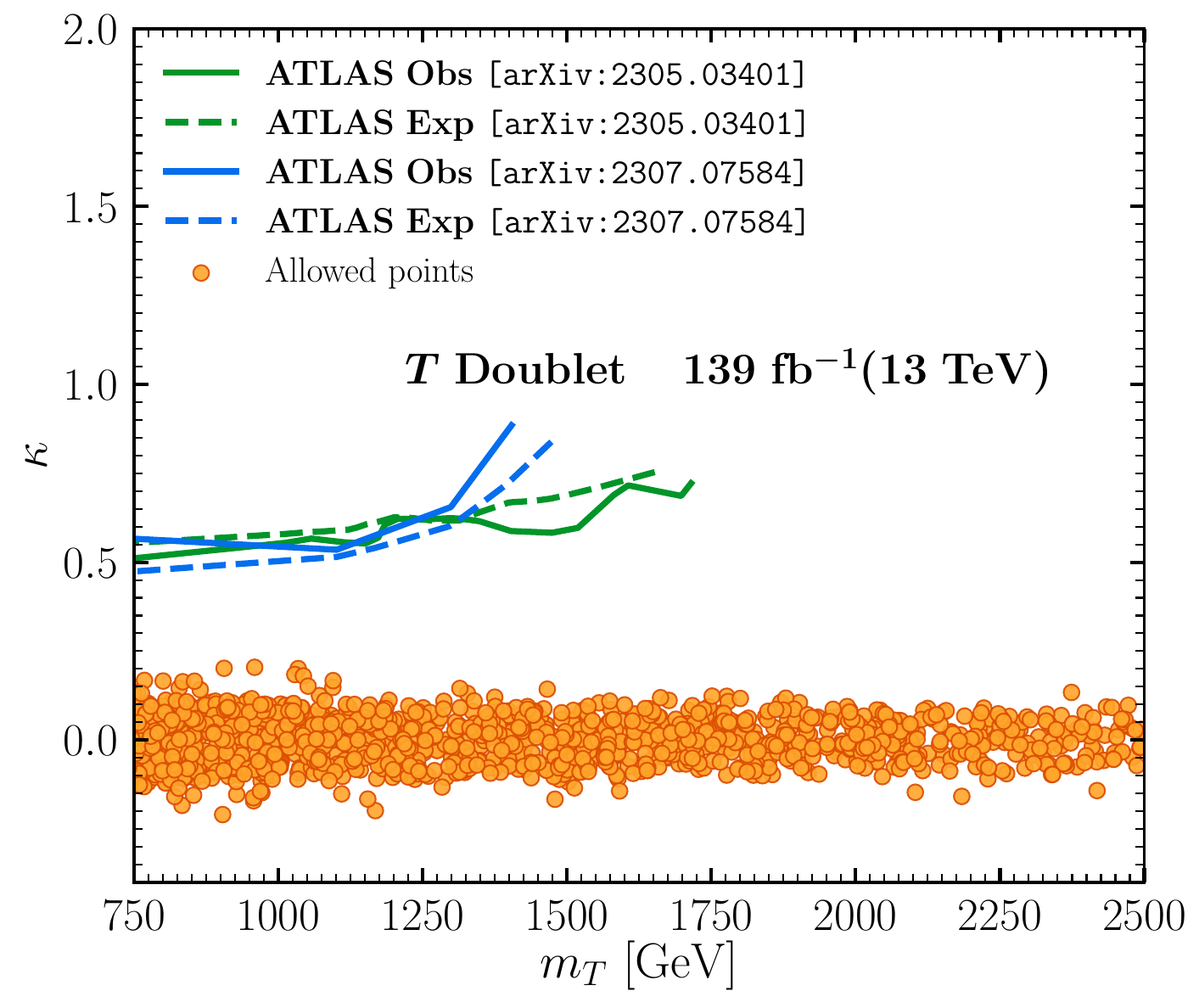}
	
	\caption{Allowed points in the ($m_T, \kappa$) plane for $(T)$ singlet left and $(TB)$ doublet right. Blue lines represent ATLAS  limits\cite{ATLAS:2023pja}  (solid for observed, dashed for expected), while green lines represent \cite{ATLAS:2023bfh}.Notably, our analysis includes an extrapolation of these limits to expand the mass range downward to 750 GeV, considering that the original dataset commenced at $\sim$1000 GeV. }\label{fig13}
\end{figure}

In this section, we evaluate the compatibility of our results with the latest LHC constraints. Figure~\ref{fig13} presents our data as orange points, juxtaposed with the ATLAS limits, represented by blue \cite{ATLAS:2023bfh} and green \cite{ ATLAS:2023pja} lines , on the $(m_T, \kappa)$ plane. The coupling $\kappa$, explained further in \cite{Cacciapaglia:2010vn,Buchkremer:2013bha}, takes the form of $s_L^uc^u_L$ in the singlet scenario (2HDM+$T$) and $s_R^uc^u_R$ in the doublet scenario (2HDM+$TB$). Importantly, , as discussed previously, the oblique parameters $S$ and $T$ impose stringent constraints on the mixing angles, limiting them to small values. This confinement is pivotal, ensuring our scenarios comply with the existing LHC limits and simultaneously allowing exploration of a broader VLQ mass ($m_T$) range.

\section{VLQ Contributions to the Oblique Parameters}\label{appSTU}
Herein, we derive and present analytical and general expressions for the two VLQ representations\footnote{We have ensured that the oblique parameters $S$, $T$, and $U$ are UV finite and independent of the renormalization scale $\mu$}, $T$ and $TB$. These expressions are formulated as functions of the scalar Passarino-Veltman functions, which are a standard in this area of study. The oblique parameters are expressed in terms of the two-point functions of the gauge bosons, as detailed below:

\begin{eqnarray}
S &=& \frac{4 s_W^2 c_W^2 }{\alpha} \Re e \Bigg[ \frac{\Pi_{\gamma \gamma}(m_Z^2)}{m_Z^2}-\frac{\Pi_{Z Z}(m_Z^2)-\Pi_{Z Z}(0)}{m_Z^2} -\Big( \frac{c_W^2-s_W^2}{c_W s_W}\Big)\Big( \frac{\Pi_{Z \gamma}(m_Z^2)+\Pi_{Z \gamma}(0)}{m_Z^2}\Big)  \Bigg] \nonumber \\ \label{eq:S_expression} \\
T &=& \frac{1}{\alpha}\Re e \left[ \frac{\Pi_{Z Z}(0)}{m_Z^2}-\frac{\Pi_{W W}(0)}{m_W^2}-\frac{2 s_W }{c_W} \frac{\Pi_{Z \gamma}(0)}{m_Z^2}\label{eq:T_expression}  \right] \\
U &=& \frac{4 s_W^2 }{\alpha} \Re e \Bigg[ \frac{\Pi_{W W}(0)-\Pi_{W W}(m_W^2)}{m_W^2} - c_W^2 \frac{\Pi_{ZZ}(0)-\Pi_{ZZ }(m_Z^2)}{m_Z^2} -2 s_W c_W \frac{\Pi_{\gamma Z}(0)-\Pi_{\gamma Z}(m_Z^2)}{m_Z^2}+ \nonumber \\ && s_W^2 \frac{\Pi_{\gamma \gamma}(m_Z^2)}{m_Z^2}  \Bigg] \label{eq:U_expression}
\end{eqnarray}
Here, $\alpha$ denotes the fine-structure constant, while $c_W$ and $s_W$ represent the cosine and sine of the Weinberg angle, respectively.

For the analytical calculations of the self-energy of gauge bosons, as depicted in equations (\ref{eq:S_expression}) and (\ref{eq:T_expression}), we employ the \texttt{FeynArts/FormCalc} public codes. These calculations specifically address scenarios where VLQs interact exclusively with third-generation SM quarks. We concentrate on the contributions from these heavy quarks, namely the SM top and bottom quarks, along with the newly introduced VLQs, since the other contributions remain the same as in the SM. With this focus, the expressions for the electroweak gauge boson self-energies can be formulated as follows:

\begin{eqnarray}
-\Pi_{\gamma Z}(q^2) &=& \sum_{f}^{}{g_{\gamma z}(c_{Zff}^L,c_{Zff}^R,Q_f,m_f^2,q^2)} \\
-\Pi_{\gamma \gamma}(q^2) &=& \sum_{f}^{}{g_{VV}(Q_f,Q_f,m_f^2,m_f^2,q^2)}\\
-\Pi_{Z Z}(q^2) &=&  2 \sum_{f_i \neq f_j}^{}{\delta_0{(|Q_{f_i}-Q_{f_j}|)} g_{VV}\left(c_{Zf_if_j}^L,c_{Zf_if_j}^R,m_{f_i}^2,m_{f_j}^2,q^2\right)}\notag\\ && +\sum_{f}^{}{g_{VV}\left(c_{Zff}^L,c_{Zff}^R,m_f^2,m_f^2,q^2\right)} \\
-\Pi_{W W}(q^2) &=&   \sum_{f_i \neq f_j}^{}{\delta_1{( |Q_{f_i}-Q_{f_j}|)} g_{VV}\left(c_{Wf_if_j}^L,c_{Wf_if_j}^R,m_{f_i}^2,m_{f_j}^2,q^2\right)}
\end{eqnarray}

Here, $c_{Vff}$ represents the couplings between the gauge boson $Z$ or $W$ and two fermions ($f=t,b,T,B$), detailed in the next subsection~\ref{app:coupling}. The functions $g_{ab}$, including $g_{VV}$ and $g_{Z\gamma}$, are defined as:

\begin{eqnarray}
g_{VV}\left(x,y,m_1^2,m_2^2,k^2 \right)&=&\frac{N_c}{8 \pi ^2}\Big[(x^2+y^2) \big( A_0(m_2^2) -2 B_{00}(k^2,m_1^2,m_2^2)+ k^2 B_1(k^2,m_1^2,m_2^2)\big)\nonumber \\ && +\big((x^2+y^2)m_1^2-2x y m_1m_2\big)B_0(k^2,m_1^2,m_2^2) \Big]\\
g_{\gamma z}\left(x,y,Q,m^2,k^2 \right)&=& \frac{-N_c}{8 \pi ^2}\Big[(x + y) Q \big(A_0(m^2) - 2 B_{00}(k^2, m^2, m^2) \notag \\ && + k^2 B_1(k^2, m^2, m^2)\big)\Big]
\end{eqnarray}
$N_c$ represents the color factor, set at $N_c=3$ for quarks. The functions $A_0$, $B_0$, $B_{00}$, and $B_1$ are the standard Passarino-Veltman functions.

\subsection{Gauge Interactions of VLQs}\label{app:coupling}
The introduction of VLQs modifies the neutral and charged current interactions. The couplings between these exotic quarks and the third-generation SM quarks, as well as the electroweak massive gauge bosons, are described as follows:

\begin{eqnarray}
Zq'q&=&  \frac{e}{2 s_w c_w}\gamma^\mu ( \kappa^{L}_{Zq'q} \mathbb{L}  + \kappa^{R}_{Zq'q} \mathbb{R} ),\nonumber \\
Wq'q&=&  \frac{e}{\sqrt{2} s_w }\gamma^\mu (\kappa^{L}_{Wq'q} \mathbb{L}  + \kappa^{R}_{Wq'q} \mathbb{R} )
\end{eqnarray}
Where $\kappa_{Vq'q}^{L,R}$ are the components for both left- and right-handed couplings of the $Z$ and $W$ bosons. Note that these couplings depend on the chosen representations of VLQs as follows:
\subsubsection{$(T)$ Singlet}
In the $SU(2)_L$ vector-like singlet scenario with $T$, the charged current couplings $\kappa_{Wq'q}^{L,R}$ are:
\begin{eqnarray}
\kappa_{Wtb}^L&=& \clx,\quad
\kappa_{Wtb}^R= 0, \nonumber  \\
\kappa_{WTb}^L&=& \slx , \quad
\kappa_{WTb}^R= 0.  \label{eq:SMT_Wff}
\end{eqnarray}
The neutral current couplings $\kappa^{L,R}_{Zq'q}$ are defined as:
\begin{eqnarray}
\kappa_{Ztt}^L&=& (\clx)^2 -\frac{4}{3} s_W^2 ,\quad
\kappa_{Ztt}^R= -\frac{4}{3} s_W^2, \nonumber  \\
\kappa_{Zbb}^L&=& -1+\frac{2}{3} s_W^2,\quad
\kappa_{Zbb}^R= \frac{2}{3} s_W^2, \nonumber  \\
\kappa_{ZTT}^L&=& (\slx)^2-\frac{4}{3} s_W^2 ,\quad
\kappa_{ZTT}^R=-\frac{4}{3} s_W^2 , \nonumber  \\
\kappa_{ZtT}^L&=& \slx \clx ,\quad
\kappa_{ZtT}^R=  0 .
\label{eq:SMT_Zff}
\end{eqnarray}

\subsubsection{$(TB)$ Doublet}
In the $(TB)$ doublet scenario, the couplings of $T$, $B$, and the third-generation SM quarks with the $W$-boson are as follows:
\begin{eqnarray}
\kappa_{Wtb}^L&=& \clu \cld + \slu \sld,\quad
\kappa_{Wtb}^R= \sru \srd,  \nonumber  \\
\kappa_{WTB}^L&=&  \clu \cld + \slu \sld ,\quad
\kappa_{WTB}^R= \cru \crd, \nonumber  \\
\kappa_{WTb}^L&=&  \slu \cld  - \clu \sld  ,\quad
\kappa_{WTb}^R= -\cru \srd, \nonumber  \\
\kappa_{WtB}^L&=& \clu \sld -\slu \cld ,\quad
\kappa_{WtB}^R= -\sru \crd.  \label{eq:SMTB_Wff}
\end{eqnarray}
The neutral current couplings to the $Z$ boson are expressed as:
\begin{eqnarray}
\kappa_{Ztt}^L&=& 1-\frac{4}{3} s_W^2,\quad
\kappa_{Ztt}^R= (\sru)^2-\frac{4}{3} s_W^2,\nonumber  \\
\kappa_{Zbb}^L&=& -1+\frac{2}{3} s_W^2,\quad
\kappa_{Zbb}^R= -(\srd)^2+\frac{2}{3} s_W^2, \nonumber  \\
\kappa_{ZTT}^L&=& 1-\frac{4}{3} s_W^2  ,\quad
\kappa_{ZTT}^R=  (\cru)^2-\frac{4}{3} s_W^2, \nonumber  \\
\kappa_{ZBB}^L&=& -1+\frac{2}{3} s_W^2 ,\quad
\kappa_{ZBB}^R= -(\crd)^2+\frac{2}{3} s_W^2, \nonumber  \\
\kappa_{ZtT}^L&=& 0 ,\quad \quad
\kappa_{ZtT}^R= -\sru \cru , \nonumber  \\
\kappa_{ZbB}^L&=& 0  ,\quad \quad
\kappa_{ZbB}^R= \srd \crd  .
\label{eq:SMTB_Zff}
\end{eqnarray}

\subsection{Passarino-Veltman Functions}
The Passarino-Veltman functions are defined as follows:

\begin{eqnarray}
A_0(m_1^2) &=& \frac{(2 \pi \mu)^{4-D}}{i \pi^{2}}\int_{}^{}{d^Dq \frac{1}{d_1}},\\
B_0;B^{\mu};B^{\mu\nu}(p_1^2,m_1^2,m_2^2) &=& \frac{(2 \pi \mu)^{4-D}}{i \pi^{2}}\int_{}^{}{d^Dq \frac{1;q^\mu;q^\mu q^\nu}{d_1d_2}}
\end{eqnarray}

The denominators $d_i$ are defined by $d_1 = q^2 - m_1^2$ and $d_2 = (q + p_1)^2 - m_2^2$, with $\mu$ being the renormalization scale. $p_1$ represents the external momentum, and $q$ is the internal momentum which is integrated out.

The functions $B^\mu$ and $B^{\mu \nu}$ are decomposed as:

\begin{eqnarray}
B^\mu &=& p_1^\mu B_1,\quad \ B^{\mu \nu } = g^{\mu \nu} B_{00} + p_1^\mu p_1^\nu B_{11}.
\end{eqnarray}

The functions $B_{00}$ and $B_1$ can be calculated by contracting $B^\mu$ and $B^{\mu\nu}$ with $p_1$ and $g_{\mu \nu}$, respectively:
\begin{eqnarray}
B_1(p^2, m_1^2, m_2^2) &=& \frac{1}{2}\left[
A_0(m_1^2) - A_0(m_2^2) - \left(p^2 + m_1^2 - m_2^2\right) B_0(p^2, m_1^2, m_2^2) \right]\\
B_{00}(p^2, m_1^2, m_2^2) &=& -\frac{
	p^2 - 3 (m_1^2 + m_2^2)}{18} + \frac{m_1^2 B_0(p^2, m_1^2, m_2^2)}{3}\notag \\ && +
\frac{A_0(m_2^2) + (p^2 + m_1^2 - m_2^2) B_1(p^2, m_1^2, m_2^2)}{6}.
\end{eqnarray}

\bibliography{ref} 

\providecommand{\href}[2]{#2}\begingroup\raggedright\begin{thebibliography}{10}

\bibitem{Davidson:1987tr}
A.~Davidson and K.C.~Wali, \emph{{Family Mass Hierarchy From Universal Seesaw
  Mechanism}}, \href{https://doi.org/10.1103/PhysRevLett.60.1813}{\emph{Phys.
  Rev. Lett.} {\bfseries 60} (1988) 1813}.

\bibitem{Babu:1989rb}
K.S.~Babu and R.N.~Mohapatra, \emph{{A Solution to the Strong {CP} Problem
  Without an Axion}},
  \href{https://doi.org/10.1103/PhysRevD.41.1286}{\emph{Phys. Rev. D}
  {\bfseries 41} (1990) 1286}.

\bibitem{Grinstein:2010ve}
B.~Grinstein, M.~Redi and G.~Villadoro, \emph{{Low Scale Flavor Gauge
  Symmetries}}, \href{https://doi.org/10.1007/JHEP11(2010)067}{\emph{JHEP}
  {\bfseries 11} (2010) 067} [\href{https://arxiv.org/abs/1009.2049}{{\ttfamily
  1009.2049}}].

\bibitem{Guadagnoli:2011id}
D.~Guadagnoli, R.N.~Mohapatra and I.~Sung, \emph{{Gauged Flavor Group with
  Left-Right Symmetry}},
  \href{https://doi.org/10.1007/JHEP04(2011)093}{\emph{JHEP} {\bfseries 04}
  (2011) 093} [\href{https://arxiv.org/abs/1103.4170}{{\ttfamily 1103.4170}}].

\bibitem{Moroi:1991mg}
T.~Moroi and Y.~Okada, \emph{{Radiative corrections to Higgs masses in the
  supersymmetric model with an extra family and antifamily}},
  \href{https://doi.org/10.1142/S0217732392000124}{\emph{Mod. Phys. Lett. A}
  {\bfseries 7} (1992) 187}.

\bibitem{Moroi:1992zk}
T.~Moroi and Y.~Okada, \emph{{Upper bound of the lightest neutral Higgs mass in
  extended supersymmetric Standard Models}},
  \href{https://doi.org/10.1016/0370-2693(92)90091-H}{\emph{Phys. Lett. B}
  {\bfseries 295} (1992) 73}.

\bibitem{Babu:2008ge}
K.S.~Babu, I.~Gogoladze, M.U.~Rehman and Q.~Shafi, \emph{{Higgs Boson Mass,
  Sparticle Spectrum and Little Hierarchy Problem in Extended MSSM}},
  \href{https://doi.org/10.1103/PhysRevD.78.055017}{\emph{Phys. Rev. D}
  {\bfseries 78} (2008) 055017}
  [\href{https://arxiv.org/abs/0807.3055}{{\ttfamily 0807.3055}}].

\bibitem{Martin:2009bg}
S.P.~Martin, \emph{{Extra vector-like matter and the lightest Higgs scalar
  boson mass in low-energy supersymmetry}},
  \href{https://doi.org/10.1103/PhysRevD.81.035004}{\emph{Phys. Rev. D}
  {\bfseries 81} (2010) 035004}
  [\href{https://arxiv.org/abs/0910.2732}{{\ttfamily 0910.2732}}].

\bibitem{Graham:2009gy}
P.W.~Graham, A.~Ismail, S.~Rajendran and P.~Saraswat, \emph{{A Little Solution
  to the Little Hierarchy Problem: A Vector-like Generation}},
  \href{https://doi.org/10.1103/PhysRevD.81.055016}{\emph{Phys. Rev. D}
  {\bfseries 81} (2010) 055016}
  [\href{https://arxiv.org/abs/0910.3020}{{\ttfamily 0910.3020}}].

\bibitem{Martin:2010dc}
S.P.~Martin, \emph{{Raising the Higgs Mass with Yukawa Couplings for
  Isotriplets in Vector-Like Extensions of Minimal Supersymmetry}},
  \href{https://doi.org/10.1103/PhysRevD.82.055019}{\emph{Phys. Rev. D}
  {\bfseries 82} (2010) 055019}
  [\href{https://arxiv.org/abs/1006.4186}{{\ttfamily 1006.4186}}].

\bibitem{Rosner:1985hx}
J.L.~Rosner, \emph{{E$_{6}$ and Exotic Fermions}}, {\emph{Comments Nucl. Part.
  Phys.} {\bfseries 15} (1986) 195}.

\bibitem{Robinett:1985dz}
R.W.~Robinett, \emph{{On the Mixing and Production of Exotic Fermions in E6}},
  \href{https://doi.org/10.1103/PhysRevD.33.1908}{\emph{Phys. Rev. D}
  {\bfseries 33} (1986) 1908}.

\bibitem{Arkani-Hamed:2002ikv}
N.~Arkani-Hamed, A.G.~Cohen, E.~Katz and A.E.~Nelson, \emph{{The Littlest
  Higgs}}, \href{https://doi.org/10.1088/1126-6708/2002/07/034}{\emph{JHEP}
  {\bfseries 07} (2002) 034}
  [\href{https://arxiv.org/abs/hep-ph/0206021}{{\ttfamily hep-ph/0206021}}].

\bibitem{Schmaltz:2005ky}
M.~Schmaltz and D.~Tucker-Smith, \emph{{Little Higgs review}},
  \href{https://doi.org/10.1146/annurev.nucl.55.090704.151502}{\emph{Ann. Rev.
  Nucl. Part. Sci.} {\bfseries 55} (2005) 229}
  [\href{https://arxiv.org/abs/hep-ph/0502182}{{\ttfamily hep-ph/0502182}}].

\bibitem{Dobrescu:1997nm}
B.A.~Dobrescu and C.T.~Hill, \emph{{Electroweak symmetry breaking via top
  condensation seesaw}},
  \href{https://doi.org/10.1103/PhysRevLett.81.2634}{\emph{Phys. Rev. Lett.}
  {\bfseries 81} (1998) 2634}
  [\href{https://arxiv.org/abs/hep-ph/9712319}{{\ttfamily hep-ph/9712319}}].

\bibitem{Chivukula:1998wd}
R.S.~Chivukula, B.A.~Dobrescu, H.~Georgi and C.T.~Hill, \emph{{Top Quark Seesaw
  Theory of Electroweak Symmetry Breaking}},
  \href{https://doi.org/10.1103/PhysRevD.59.075003}{\emph{Phys. Rev. D}
  {\bfseries 59} (1999) 075003}
  [\href{https://arxiv.org/abs/hep-ph/9809470}{{\ttfamily hep-ph/9809470}}].

\bibitem{He:2001fz}
H.-J.~He, C.T.~Hill and T.M.P.~Tait, \emph{{Top Quark Seesaw, Vacuum Structure
  and Electroweak Precision Constraints}},
  \href{https://doi.org/10.1103/PhysRevD.65.055006}{\emph{Phys. Rev. D}
  {\bfseries 65} (2002) 055006}
  [\href{https://arxiv.org/abs/hep-ph/0108041}{{\ttfamily hep-ph/0108041}}].

\bibitem{Hill:2002ap}
C.T.~Hill and E.H.~Simmons, \emph{{Strong Dynamics and Electroweak Symmetry
  Breaking}}, \href{https://doi.org/10.1016/S0370-1573(03)00140-6}{\emph{Phys.
  Rept.} {\bfseries 381} (2003) 235}
  [\href{https://arxiv.org/abs/hep-ph/0203079}{{\ttfamily hep-ph/0203079}}].

\bibitem{Agashe:2004rs}
K.~Agashe, R.~Contino and A.~Pomarol, \emph{{The Minimal composite Higgs
  model}}, \href{https://doi.org/10.1016/j.nuclphysb.2005.04.035}{\emph{Nucl.
  Phys. B} {\bfseries 719} (2005) 165}
  [\href{https://arxiv.org/abs/hep-ph/0412089}{{\ttfamily hep-ph/0412089}}].

\bibitem{Contino:2006qr}
R.~Contino, L.~Da~Rold and A.~Pomarol, \emph{{Light custodians in natural
  composite Higgs models}},
  \href{https://doi.org/10.1103/PhysRevD.75.055014}{\emph{Phys. Rev. D}
  {\bfseries 75} (2007) 055014}
  [\href{https://arxiv.org/abs/hep-ph/0612048}{{\ttfamily hep-ph/0612048}}].

\bibitem{Barbieri:2007bh}
R.~Barbieri, B.~Bellazzini, V.S.~Rychkov and A.~Varagnolo, \emph{{The Higgs
  boson from an extended symmetry}},
  \href{https://doi.org/10.1103/PhysRevD.76.115008}{\emph{Phys. Rev. D}
  {\bfseries 76} (2007) 115008}
  [\href{https://arxiv.org/abs/0706.0432}{{\ttfamily 0706.0432}}].

\bibitem{Anastasiou:2009rv}
C.~Anastasiou, E.~Furlan and J.~Santiago, \emph{{Realistic Composite Higgs
  Models}}, \href{https://doi.org/10.1103/PhysRevD.79.075003}{\emph{Phys. Rev.
  D} {\bfseries 79} (2009) 075003}
  [\href{https://arxiv.org/abs/0901.2117}{{\ttfamily 0901.2117}}].

\bibitem{Aguilar-Saavedra:2009xmz}
J.A.~Aguilar-Saavedra, \emph{{Identifying top partners at LHC}},
  \href{https://doi.org/10.1088/1126-6708/2009/11/030}{\emph{JHEP} {\bfseries
  11} (2009) 030} [\href{https://arxiv.org/abs/0907.3155}{{\ttfamily
  0907.3155}}].

\bibitem{Okada:2012gy}
Y.~Okada and L.~Panizzi, \emph{{LHC signatures of vector-like quarks}},
  \href{https://doi.org/10.1155/2013/364936}{\emph{Adv. High Energy Phys.}
  {\bfseries 2013} (2013) 364936}
  [\href{https://arxiv.org/abs/1207.5607}{{\ttfamily 1207.5607}}].

\bibitem{DeSimone:2012fs}
A.~De~Simone, O.~Matsedonskyi, R.~Rattazzi and A.~Wulzer, \emph{{A First Top
  Partner Hunter's Guide}},
  \href{https://doi.org/10.1007/JHEP04(2013)004}{\emph{JHEP} {\bfseries 04}
  (2013) 004} [\href{https://arxiv.org/abs/1211.5663}{{\ttfamily 1211.5663}}].

\bibitem{Buchkremer:2013bha}
M.~Buchkremer, G.~Cacciapaglia, A.~Deandrea and L.~Panizzi, \emph{{Model
  Independent Framework for Searches of Top Partners}},
  \href{https://doi.org/10.1016/j.nuclphysb.2013.08.010}{\emph{Nucl. Phys. B}
  {\bfseries 876} (2013) 376}
  [\href{https://arxiv.org/abs/1305.4172}{{\ttfamily 1305.4172}}].

\bibitem{Aguilar-Saavedra:2013qpa}
J.A.~Aguilar-Saavedra, R.~Benbrik, S.~Heinemeyer and M.~P\'erez-Victoria,
  \emph{{Handbook of vectorlike quarks: Mixing and single production}},
  \href{https://doi.org/10.1103/PhysRevD.88.094010}{\emph{Phys. Rev. D}
  {\bfseries 88} (2013) 094010}
  [\href{https://arxiv.org/abs/1306.0572}{{\ttfamily 1306.0572}}].

\bibitem{ATLAS:2017vdo}
{\scshape \bf ATLAS} collaboration, \emph{{Search for pair production of
  vector-like top quarks in events with one lepton, jets, and missing
  transverse momentum in $ \sqrt{s}=13 $ TeV $pp$ collisions with the ATLAS
  detector}}, \href{https://doi.org/10.1007/JHEP08(2017)052}{\emph{JHEP}
  {\bfseries 08} (2017) 052}
  [\href{https://arxiv.org/abs/1705.10751}{{\ttfamily 1705.10751}}].

\bibitem{ATLAS:2017nap}
{\scshape \bf ATLAS} collaboration, \emph{{Search for pair production of heavy
  vector-like quarks decaying to high-p$_{T}$ W bosons and b quarks in the
  lepton-plus-jets final state in pp collisions at $ \sqrt{s}=13 $ TeV with the
  ATLAS detector}}, \href{https://doi.org/10.1007/JHEP10(2017)141}{\emph{JHEP}
  {\bfseries 10} (2017) 141}
  [\href{https://arxiv.org/abs/1707.03347}{{\ttfamily 1707.03347}}].

\bibitem{ATLAS:2018cye}
{\scshape \bf ATLAS} collaboration, \emph{{Search for pair production of
  up-type vector-like quarks and for four-top-quark events in final states with
  multiple $b$-jets with the ATLAS detector}},
  \href{https://doi.org/10.1007/JHEP07(2018)089}{\emph{JHEP} {\bfseries 07}
  (2018) 089} [\href{https://arxiv.org/abs/1803.09678}{{\ttfamily
  1803.09678}}].

\bibitem{CMS:2017ked}
{\scshape \bf CMS} collaboration, \emph{{Search for pair production of
  vector-like T and B quarks in single-lepton final states using boosted jet
  substructure in proton-proton collisions at $\sqrt{s}=13$ TeV}},
  \href{https://doi.org/10.1007/JHEP11(2017)085}{\emph{JHEP} {\bfseries 11}
  (2017) 085} [\href{https://arxiv.org/abs/1706.03408}{{\ttfamily
  1706.03408}}].

\bibitem{CMS:2017ynm}
{\scshape \bf CMS} collaboration, \emph{{Search for pair production of
  vector-like quarks in the bW$\overline{\mathrm{b}}$W channel from
  proton-proton collisions at $\sqrt{s} =$ 13 TeV}},
  \href{https://doi.org/10.1016/j.physletb.2018.01.077}{\emph{Phys. Lett. B}
  {\bfseries 779} (2018) 82}
  [\href{https://arxiv.org/abs/1710.01539}{{\ttfamily 1710.01539}}].

\bibitem{CMS:2018zkf}
{\scshape \bf CMS} collaboration, \emph{{Search for vector-like T and B quark
  pairs in final states with leptons at $\sqrt{s} =$ 13 TeV}},
  \href{https://doi.org/10.1007/JHEP08(2018)177}{\emph{JHEP} {\bfseries 08}
  (2018) 177} [\href{https://arxiv.org/abs/1805.04758}{{\ttfamily
  1805.04758}}].

\bibitem{CMS:2017voh}
{\scshape \bf CMS} collaboration, \emph{{Search for single production of a
  vector-like T quark decaying to a Z boson and a top quark in proton-proton
  collisions at $\sqrt s$ = 13 TeV}},
  \href{https://doi.org/10.1016/j.physletb.2018.04.036}{\emph{Phys. Lett. B}
  {\bfseries 781} (2018) 574}
  [\href{https://arxiv.org/abs/1708.01062}{{\ttfamily 1708.01062}}].

\bibitem{CMS:2018kcw}
{\scshape \bf CMS} collaboration, \emph{{Search for single production of
  vector-like quarks decaying to a b quark and a Higgs boson}},
  \href{https://doi.org/10.1007/JHEP06(2018)031}{\emph{JHEP} {\bfseries 06}
  (2018) 031} [\href{https://arxiv.org/abs/1802.01486}{{\ttfamily
  1802.01486}}].

\bibitem{ATLAS:2016ovj}
{\scshape \bf ATLAS} collaboration, \emph{{Search for single production of
  vector-like quarks decaying into $Wb$ in $pp$ collisions at $\sqrt{s} =$ 13
  TeV with the ATLAS detector}}, .

\bibitem{ATLAS:2018mpo}
{\scshape \bf ATLAS} collaboration, \emph{{Search for pair production of heavy
  vector-like quarks decaying into high-$p_T$ $W$ bosons and top quarks in the
  lepton-plus-jets final state in $pp$ collisions at $\sqrt{s}=13$ TeV with the
  ATLAS detector}}, \href{https://doi.org/10.1007/JHEP08(2018)048}{\emph{JHEP}
  {\bfseries 08} (2018) 048}
  [\href{https://arxiv.org/abs/1806.01762}{{\ttfamily 1806.01762}}].

\bibitem{ATLAS:2018uky}
{\scshape \bf ATLAS} collaboration, \emph{{Search for pair production of heavy
  vector-like quarks decaying into hadronic final states in $pp$ collisions at
  $\sqrt{s} = 13$ TeV with the ATLAS detector}},
  \href{https://doi.org/10.1103/PhysRevD.98.092005}{\emph{Phys. Rev. D}
  {\bfseries 98} (2018) 092005}
  [\href{https://arxiv.org/abs/1808.01771}{{\ttfamily 1808.01771}}].

\bibitem{ATLAS:2018ziw}
{\scshape \bf ATLAS} collaboration, \emph{{Combination of the searches for
  pair-produced vector-like partners of the third-generation quarks at
  $\sqrt{s} =$ 13 TeV with the ATLAS detector}},
  \href{https://doi.org/10.1103/PhysRevLett.121.211801}{\emph{Phys. Rev. Lett.}
  {\bfseries 121} (2018) 211801}
  [\href{https://arxiv.org/abs/1808.02343}{{\ttfamily 1808.02343}}].

\bibitem{ATLAS:2018dyh}
{\scshape \bf ATLAS} collaboration, \emph{{Search for single production of
  vector-like quarks decaying into $Wb$ in $pp$ collisions at $\sqrt{s} = 13$
  TeV with the ATLAS detector}},
  \href{https://doi.org/10.1007/JHEP05(2019)164}{\emph{JHEP} {\bfseries 05}
  (2019) 164} [\href{https://arxiv.org/abs/1812.07343}{{\ttfamily
  1812.07343}}].

\bibitem{ATLAS:2021ddx}
{\scshape \bf ATLAS} collaboration, \emph{{Search for single production of
  vector-like $T$ quarks decaying to $Ht$ or $Zt$ in $pp$ collisions at
  $\sqrt{s}$ = 13 TeV with the ATLAS detector}}, .

\bibitem{ATLAS:2021gfv}
{\scshape \bf ATLAS} collaboration, \emph{{Search for single vector-like $B$
  quark production and decay via $B\rightarrow bH(b\bar{b})$ in pp collisions
  at $\sqrt{s} = 13\text{ TeV}$ with the ATLAS detector}}, .

\bibitem{ATLAS:2021ibc}
{\scshape \bf ATLAS} collaboration, \emph{{Search for pair-production of
  vector-like quarks in $pp$ collision events at
  $\sqrt{s}=13$\textasciitilde{}TeV with at least one leptonically-decaying
  $Z$\textasciitilde{}boson and a third-generation quark with the ATLAS
  detector}}, .

\bibitem{CMS:2016edj}
{\scshape \bf CMS} collaboration, \emph{{Search for single production of a
  heavy vector-like T quark decaying to a Higgs boson and a top quark with a
  lepton and jets in the final state}},
  \href{https://doi.org/10.1016/j.physletb.2017.05.019}{\emph{Phys. Lett. B}
  {\bfseries 771} (2017) 80}
  [\href{https://arxiv.org/abs/1612.00999}{{\ttfamily 1612.00999}}].

\bibitem{CMS:2017fpk}
{\scshape \bf CMS} collaboration, \emph{{Search for single production of
  vector-like quarks decaying into a b quark and a W boson in proton-proton
  collisions at $\sqrt s =$ 13 TeV}},
  \href{https://doi.org/10.1016/j.physletb.2017.07.022}{\emph{Phys. Lett. B}
  {\bfseries 772} (2017) 634}
  [\href{https://arxiv.org/abs/1701.08328}{{\ttfamily 1701.08328}}].

\bibitem{CMS:2018haz}
{\scshape \bf CMS} collaboration, \emph{{Search for a vector-like quark
  decaying to a top quark and a W boson}}, .

\bibitem{CMS:2018dcw}
{\scshape \bf CMS} collaboration, \emph{{Search for single production of
  vector-like quarks decaying to a top quark and a W boson in proton-proton
  collisions at $\sqrt{s} =$ 13 TeV}},
  \href{https://doi.org/10.1140/epjc/s10052-019-6556-3}{\emph{Eur. Phys. J. C}
  {\bfseries 79} (2019) 90} [\href{https://arxiv.org/abs/1809.08597}{{\ttfamily
  1809.08597}}].

\bibitem{CMS:2018rxs}
{\scshape \bf CMS} collaboration, \emph{{Search for a W' boson decaying to a
  vector-like quark and a top or bottom quark in the all-jets final state}},
  \href{https://doi.org/10.1007/JHEP03(2019)127}{\emph{JHEP} {\bfseries 03}
  (2019) 127} [\href{https://arxiv.org/abs/1811.07010}{{\ttfamily
  1811.07010}}].

\bibitem{CMS:2020fuj}
{\scshape \bf CMS} collaboration, \emph{{A search for bottom-type, vector-like
  quark pair production in a fully hadronic mode in proton-proton collisions at
  $\sqrt{s} = 13$ TeV}}, .

\bibitem{CMS:2020ttz}
{\scshape \bf CMS} collaboration, \emph{{A search for bottom-type, vector-like
  quark pair production in a fully hadronic final state in proton-proton
  collisions at $\sqrt{s} =$ 13 TeV}},
  \href{https://doi.org/10.1103/PhysRevD.102.112004}{\emph{Phys. Rev. D}
  {\bfseries 102} (2020) 112004}
  [\href{https://arxiv.org/abs/2008.09835}{{\ttfamily 2008.09835}}].

\bibitem{CMS:2022yxp}
{\scshape \bf CMS} collaboration, \emph{{Search for single production of a
  vector-like T quark decaying to a top quark and a Z boson in the final state
  with jets and missing transverse momentum at $ \sqrt{s} $ = 13 TeV}},
  \href{https://doi.org/10.1007/JHEP05(2022)093}{\emph{JHEP} {\bfseries 05}
  (2022) 093} [\href{https://arxiv.org/abs/2201.02227}{{\ttfamily
  2201.02227}}].

\bibitem{CMS:2022tdo}
{\scshape \bf CMS} collaboration, \emph{{Search for a W' boson decaying to a
  vector-like quark and a top or bottom quark in the all-jets final state at $
  \sqrt{\mathrm{s}} $ = 13 TeV}},
  \href{https://doi.org/10.1007/JHEP09(2022)088}{\emph{JHEP} {\bfseries 09}
  (2022) 088} [\href{https://arxiv.org/abs/2202.12988}{{\ttfamily
  2202.12988}}].

\bibitem{CMS:2022fck}
{\scshape \bf CMS} collaboration, \emph{{Search for pair production of
  vector-like quarks in leptonic final states in proton-proton collisions at $
  \sqrt{s} $ = 13 TeV}},
  \href{https://doi.org/10.1007/JHEP07(2023)020}{\emph{JHEP} {\bfseries 07}
  (2023) 020} [\href{https://arxiv.org/abs/2209.07327}{{\ttfamily
  2209.07327}}].

\bibitem{Benbrik:2015fyz}
R.~Benbrik, C.-H.~Chen and T.~Nomura, \emph{{Higgs singlet boson as a diphoton
  resonance in a vectorlike quark model}},
  \href{https://doi.org/10.1103/PhysRevD.93.055034}{\emph{Phys. Rev. D}
  {\bfseries 93} (2016) 055034}
  [\href{https://arxiv.org/abs/1512.06028}{{\ttfamily 1512.06028}}].

\bibitem{Badziak:2015zez}
M.~Badziak, \emph{{Interpreting the 750 GeV diphoton excess in minimal
  extensions of Two-Higgs-Doublet models}},
  \href{https://doi.org/10.1016/j.physletb.2016.06.003}{\emph{Phys. Lett. B}
  {\bfseries 759} (2016) 464}
  [\href{https://arxiv.org/abs/1512.07497}{{\ttfamily 1512.07497}}].

\bibitem{Angelescu:2015uiz}
A.~Angelescu, A.~Djouadi and G.~Moreau, \emph{{Scenarii for interpretations of
  the LHC diphoton excess: two Higgs doublets and vector-like quarks and
  leptons}}, \href{https://doi.org/10.1016/j.physletb.2016.02.064}{\emph{Phys.
  Lett. B} {\bfseries 756} (2016) 126}
  [\href{https://arxiv.org/abs/1512.04921}{{\ttfamily 1512.04921}}].

\bibitem{Arhrib:2016rlj}
A.~Arhrib, R.~Benbrik, S.J.D.~King, B.~Manaut, S.~Moretti and C.S.~Un,
  \emph{{Phenomenology of 2HDM with vectorlike quarks}},
  \href{https://doi.org/10.1103/PhysRevD.97.095015}{\emph{Phys. Rev. D}
  {\bfseries 97} (2018) 095015}
  [\href{https://arxiv.org/abs/1607.08517}{{\ttfamily 1607.08517}}].

\bibitem{Benbrik:2019zdp}
R.~Benbrik et~al., \emph{{Signatures of vector-like top partners decaying into
  new neutral scalar or pseudoscalar bosons}},
  \href{https://doi.org/10.1007/JHEP05(2020)028}{\emph{JHEP} {\bfseries 05}
  (2020) 028} [\href{https://arxiv.org/abs/1907.05929}{{\ttfamily
  1907.05929}}].

\bibitem{Branco:2011iw}
G.C.~Branco, P.M.~Ferreira, L.~Lavoura, M.N.~Rebelo, M.~Sher and J.P.~Silva,
  \emph{{Theory and phenomenology of two-Higgs-doublet models}},
  \href{https://doi.org/10.1016/j.physrep.2012.02.002}{\emph{Phys. Rept.}
  {\bfseries 516} (2012) 1} [\href{https://arxiv.org/abs/1106.0034}{{\ttfamily
  1106.0034}}].

\bibitem{Gunion:1989we}
J.F.~Gunion, H.E.~Haber, G.L.~Kane and S.~Dawson, \emph{{The Higgs Hunter's
  Guide}}, vol.~80 (2000).

\bibitem{Peskin:1991sw}
M.E.~Peskin and T.~Takeuchi, \emph{{Estimation of oblique electroweak
  corrections}}, \href{https://doi.org/10.1103/PhysRevD.46.381}{\emph{Phys.
  Rev. D} {\bfseries 46} (1992) 381}.

\bibitem{Abouabid:2023mbu}
H.~Abouabid, A.~Arhrib, R.~Benbrik, M.~Boukidi and J.E.~Falaki, \emph{{The
  oblique parameters in the 2HDM with Vector-Like Quarks: Confronting $M_W$
  CDF-II Anomaly}},  \href{https://arxiv.org/abs/2302.07149}{{\ttfamily
  2302.07149}}.

\bibitem{Altarelli:1990zd}
G.~Altarelli and R.~Barbieri, \emph{{Vacuum polarization effects of new physics
  on electroweak processes}},
  \href{https://doi.org/10.1016/0370-2693(91)91378-9}{\emph{Phys. Lett. B}
  {\bfseries 253} (1991) 161}.

\bibitem{Vignaroli:2015ama}
N.~Vignaroli, \emph{{$Z$-peaked excess from heavy gluon decays to vectorlike
  quarks}}, \href{https://doi.org/10.1103/PhysRevD.91.115009}{\emph{Phys. Rev.
  D} {\bfseries 91} (2015) 115009}
  [\href{https://arxiv.org/abs/1504.01768}{{\ttfamily 1504.01768}}].

\bibitem{Boudjema:1989qga}
F.~Boudjema, A.~Djouadi and C.~Verzegnassi, \emph{{A General Sum Rule for the
  Top Mass From $b$ Physics on $Z$ Resonance}},
  \href{https://doi.org/10.1016/0370-2693(90)91759-5}{\emph{Phys. Lett. B}
  {\bfseries 238} (1990) 423}.

\bibitem{Kanemura:1993hm}
S.~Kanemura, T.~Kubota and E.~Takasugi, \emph{{Lee-Quigg-Thacker bounds for
  Higgs boson masses in a two doublet model}},
  \href{https://doi.org/10.1016/0370-2693(93)91205-2}{\emph{Phys. Lett. B}
  {\bfseries 313} (1993) 155}
  [\href{https://arxiv.org/abs/hep-ph/9303263}{{\ttfamily hep-ph/9303263}}].

\bibitem{Barroso:2013awa}
A.~Barroso, P.M.~Ferreira, I.P.~Ivanov and R.~Santos, \emph{{Metastability
  bounds on the two Higgs doublet model}},
  \href{https://doi.org/10.1007/JHEP06(2013)045}{\emph{JHEP} {\bfseries 06}
  (2013) 045} [\href{https://arxiv.org/abs/1303.5098}{{\ttfamily 1303.5098}}].

\bibitem{Deshpande:1977rw}
N.G.~Deshpande and E.~Ma, \emph{{Pattern of Symmetry Breaking with Two Higgs
  Doublets}}, \href{https://doi.org/10.1103/PhysRevD.18.2574}{\emph{Phys. Rev.
  D} {\bfseries 18} (1978) 2574}.

\bibitem{Hahn:2000kx}
T.~Hahn, \emph{{Generating Feynman diagrams and amplitudes with FeynArts 3}},
  \href{https://doi.org/10.1016/S0010-4655(01)00290-9}{\emph{Comput. Phys.
  Commun.} {\bfseries 140} (2001) 418}
  [\href{https://arxiv.org/abs/hep-ph/0012260}{{\ttfamily hep-ph/0012260}}].

\bibitem{Hahn:2001rv}
T.~Hahn and C.~Schappacher, \emph{{The Implementation of the minimal
  supersymmetric standard model in FeynArts and FormCalc}},
  \href{https://doi.org/10.1016/S0010-4655(01)00436-2}{\emph{Comput. Phys.
  Commun.} {\bfseries 143} (2002) 54}
  [\href{https://arxiv.org/abs/hep-ph/0105349}{{\ttfamily hep-ph/0105349}}].

\bibitem{Grimus:2007if}
W.~Grimus, L.~Lavoura, O.M.~Ogreid and P.~Osland, \emph{{A Precision constraint
  on multi-Higgs-doublet models}},
  \href{https://doi.org/10.1088/0954-3899/35/7/075001}{\emph{J. Phys. G}
  {\bfseries 35} (2008) 075001}
  [\href{https://arxiv.org/abs/0711.4022}{{\ttfamily 0711.4022}}].

\bibitem{Molewski:2021ogs}
M.J.~Molewski and B.J.P.~Jones, \emph{{Scalable qubit representations of
  neutrino mixing matrices}},
  \href{https://doi.org/10.1103/PhysRevD.105.056024}{\emph{Phys. Rev. D}
  {\bfseries 105} (2022) 056024}
  [\href{https://arxiv.org/abs/2111.05401}{{\ttfamily 2111.05401}}].

\bibitem{Eriksson:2009ws}
D.~Eriksson, J.~Rathsman and O.~Stal, \emph{{2HDMC: Two-Higgs-Doublet Model
  Calculator Physics and Manual}},
  \href{https://doi.org/10.1016/j.cpc.2009.09.011}{\emph{Comput. Phys. Commun.}
  {\bfseries 181} (2010) 189}
  [\href{https://arxiv.org/abs/0902.0851}{{\ttfamily 0902.0851}}].

\bibitem{Bechtle:2020pkv}
P.~Bechtle, D.~Dercks, S.~Heinemeyer, T.~Klingl, T.~Stefaniak, G.~Weiglein
  et~al., \emph{{HiggsBounds-5: Testing Higgs Sectors in the LHC 13 TeV Era}},
  \href{https://doi.org/10.1140/epjc/s10052-020-08557-9}{\emph{Eur. Phys. J. C}
  {\bfseries 80} (2020) 1211}
  [\href{https://arxiv.org/abs/2006.06007}{{\ttfamily 2006.06007}}].

\bibitem{Bechtle:2020uwn}
P.~Bechtle, S.~Heinemeyer, T.~Klingl, T.~Stefaniak, G.~Weiglein and
  J.~Wittbrodt, \emph{{HiggsSignals-2: Probing new physics with precision Higgs
  measurements in the LHC 13 TeV era}},
  \href{https://doi.org/10.1140/epjc/s10052-021-08942-y}{\emph{Eur. Phys. J. C}
  {\bfseries 81} (2021) 145}
  [\href{https://arxiv.org/abs/2012.09197}{{\ttfamily 2012.09197}}].

\bibitem{Bahl:2022igd}
H.~Bahl, T.~Biek\"otter, S.~Heinemeyer, C.~Li, S.~Paasch, G.~Weiglein et~al.,
  \emph{{HiggsTools: BSM scalar phenomenology with new versions of HiggsBounds
  and HiggsSignals}},
  \href{https://doi.org/10.1016/j.cpc.2023.108803}{\emph{Comput. Phys. Commun.}
  {\bfseries 291} (2023) 108803}
  [\href{https://arxiv.org/abs/2210.09332}{{\ttfamily 2210.09332}}].

\bibitem{Bechtle:2008jh}
P.~Bechtle, O.~Brein, S.~Heinemeyer, G.~Weiglein and K.E.~Williams,
  \emph{{HiggsBounds: Confronting Arbitrary Higgs Sectors with Exclusion Bounds
  from LEP and the Tevatron}},
  \href{https://doi.org/10.1016/j.cpc.2009.09.003}{\emph{Comput. Phys. Commun.}
  {\bfseries 181} (2010) 138}
  [\href{https://arxiv.org/abs/0811.4169}{{\ttfamily 0811.4169}}].

\bibitem{Bechtle:2011sb}
P.~Bechtle, O.~Brein, S.~Heinemeyer, G.~Weiglein and K.E.~Williams,
  \emph{{HiggsBounds 2.0.0: Confronting Neutral and Charged Higgs Sector
  Predictions with Exclusion Bounds from LEP and the Tevatron}},
  \href{https://doi.org/10.1016/j.cpc.2011.07.015}{\emph{Comput. Phys. Commun.}
  {\bfseries 182} (2011) 2605}
  [\href{https://arxiv.org/abs/1102.1898}{{\ttfamily 1102.1898}}].

\bibitem{Bechtle:2013wla}
P.~Bechtle, O.~Brein, S.~Heinemeyer, O.~St\r{a}l, T.~Stefaniak, G.~Weiglein
  et~al., \emph{{$\mathsf{HiggsBounds}-4$: Improved Tests of Extended Higgs
  Sectors against Exclusion Bounds from LEP, the Tevatron and the LHC}},
  \href{https://doi.org/10.1140/epjc/s10052-013-2693-2}{\emph{Eur. Phys. J. C}
  {\bfseries 74} (2014) 2693}
  [\href{https://arxiv.org/abs/1311.0055}{{\ttfamily 1311.0055}}].

\bibitem{Bechtle:2015pma}
P.~Bechtle, S.~Heinemeyer, O.~Stal, T.~Stefaniak and G.~Weiglein,
  \emph{{Applying Exclusion Likelihoods from LHC Searches to Extended Higgs
  Sectors}}, \href{https://doi.org/10.1140/epjc/s10052-015-3650-z}{\emph{Eur.
  Phys. J. C} {\bfseries 75} (2015) 421}
  [\href{https://arxiv.org/abs/1507.06706}{{\ttfamily 1507.06706}}].

\bibitem{Mahmoudi:2008tp}
F.~Mahmoudi, \emph{{SuperIso v2.3: A Program for calculating flavor physics
  observables in Supersymmetry}},
  \href{https://doi.org/10.1016/j.cpc.2009.02.017}{\emph{Comput. Phys. Commun.}
  {\bfseries 180} (2009) 1579}
  [\href{https://arxiv.org/abs/0808.3144}{{\ttfamily 0808.3144}}].

\bibitem{HFLAV:2016hnz}
{\scshape \bf HFLAV} collaboration, \emph{{Averages of $b$-hadron, $c$-hadron,
  and $\tau$-lepton properties as of summer 2016}},
  \href{https://doi.org/10.1140/epjc/s10052-017-5058-4}{\emph{Eur. Phys. J. C}
  {\bfseries 77} (2017) 895}
  [\href{https://arxiv.org/abs/1612.07233}{{\ttfamily 1612.07233}}].

\bibitem{CMS:2022mgd}
{\scshape \bf CMS} collaboration, \emph{{Measurement of the
  Bs0\textrightarrow{}\ensuremath{\mu}+\ensuremath{\mu}\ensuremath{-} decay
  properties and search for the
  B0\textrightarrow{}\ensuremath{\mu}+\ensuremath{\mu}\ensuremath{-} decay in
  proton-proton collisions at s=13TeV}},
  \href{https://doi.org/10.1016/j.physletb.2023.137955}{\emph{Phys. Lett. B}
  {\bfseries 842} (2023) 137955}
  [\href{https://arxiv.org/abs/2212.10311}{{\ttfamily 2212.10311}}].

\bibitem{LHCb:2021awg}
{\scshape \bf LHCb} collaboration, \emph{{Measurement of the
  $B^0_s\to\mu^+\mu^-$ decay properties and search for the $B^0\to\mu^+\mu^-$
  and $B^0_s\to\mu^+\mu^-\gamma$ decays}},
  \href{https://doi.org/10.1103/PhysRevD.105.012010}{\emph{Phys. Rev. D}
  {\bfseries 105} (2022) 012010}
  [\href{https://arxiv.org/abs/2108.09283}{{\ttfamily 2108.09283}}].

\bibitem{LHCb:2021vsc}
{\scshape \bf LHCb} collaboration, \emph{{Analysis of Neutral B-Meson Decays
  into Two Muons}},
  \href{https://doi.org/10.1103/PhysRevLett.128.041801}{\emph{Phys. Rev. Lett.}
  {\bfseries 128} (2022) 041801}
  [\href{https://arxiv.org/abs/2108.09284}{{\ttfamily 2108.09284}}].

\bibitem{Hermann:2012fc}
T.~Hermann, M.~Misiak and M.~Steinhauser, \emph{{$\bar{B}\to X_s \gamma$ in the
  Two Higgs Doublet Model up to Next-to-Next-to-Leading Order in QCD}},
  \href{https://doi.org/10.1007/JHEP11(2012)036}{\emph{JHEP} {\bfseries 11}
  (2012) 036} [\href{https://arxiv.org/abs/1208.2788}{{\ttfamily 1208.2788}}].

\bibitem{Alwall:2014hca}
J.~Alwall, R.~Frederix, S.~Frixione, V.~Hirschi, F.~Maltoni, O.~Mattelaer
  et~al., \emph{{The automated computation of tree-level and next-to-leading
  order differential cross sections, and their matching to parton shower
  simulations}}, \href{https://doi.org/10.1007/JHEP07(2014)079}{\emph{JHEP}
  {\bfseries 07} (2014) 079} [\href{https://arxiv.org/abs/1405.0301}{{\ttfamily
  1405.0301}}].

\bibitem{Pumplin:2002vw}
J.~Pumplin, D.R.~Stump, J.~Huston, H.L.~Lai, P.M.~Nadolsky and W.K.~Tung,
  \emph{{New generation of parton distributions with uncertainties from global
  QCD analysis}},
  \href{https://doi.org/10.1088/1126-6708/2002/07/012}{\emph{JHEP} {\bfseries
  07} (2002) 012} [\href{https://arxiv.org/abs/hep-ph/0201195}{{\ttfamily
  hep-ph/0201195}}].

\bibitem{Dermisek:2020gbr}
R.~Dermisek, E.~Lunghi, N.~McGinnis and S.~Shin, \emph{{Signals with six bottom
  quarks for charged and neutral Higgs bosons}},
  \href{https://doi.org/10.1007/JHEP07(2020)241}{\emph{JHEP} {\bfseries 07}
  (2020) 241} [\href{https://arxiv.org/abs/2005.07222}{{\ttfamily
  2005.07222}}].

\bibitem{ATLAS:2023bfh}
{\scshape \bf ATLAS} collaboration, \emph{{Search for singly produced
  vector-like top partners in multilepton final states with 139
  $\mathrm{fb}^{-1}$ of $pp$ collision data at $\sqrt{s} = 13$ TeV with the
  ATLAS detector}},  \href{https://arxiv.org/abs/2307.07584}{{\ttfamily
  2307.07584}}.

\bibitem{ATLAS:2023pja}
{\scshape \bf ATLAS} collaboration, \emph{{Search for single production of
  vector-like T quarks decaying into Ht or Zt in pp collisions at $ \sqrt{s} $
  = 13 TeV with the ATLAS detector}},
  \href{https://doi.org/10.1007/JHEP08(2023)153}{\emph{JHEP} {\bfseries 08}
  (2023) 153} [\href{https://arxiv.org/abs/2305.03401}{{\ttfamily
  2305.03401}}].

\bibitem{Cacciapaglia:2010vn}
G.~Cacciapaglia, A.~Deandrea, D.~Harada and Y.~Okada, \emph{{Bounds and Decays
  of New Heavy Vector-like Top Partners}},
  \href{https://doi.org/10.1007/JHEP11(2010)159}{\emph{JHEP} {\bfseries 11}
  (2010) 159} [\href{https://arxiv.org/abs/1007.2933}{{\ttfamily 1007.2933}}].

\end{thebibliography}\endgroup
\bibliographystyle{JHEP}

	\end{document}